\documentclass{aastex}
\usepackage{spr-astr-addons}
\usepackage{url}

\RequirePackage{color}

\usepackage{graphicx}
\usepackage[latin1]{inputenc}
\usepackage[T1]{fontenc}
\usepackage[english]{babel}
\usepackage{epsfig}
\usepackage{graphicx}
\usepackage{amsmath}
\usepackage{pstricks}
\usepackage{subfigure}
\usepackage{txfonts}
\begin{document}

\title{Viscous $\Lambda \textrm{CDM}$ Universe Models}

\author{Nouraddin Mostafapoor} \and \author{Øyvind Grøn\altaffilmark{1,2}}

\altaffiltext{1}{Department of Physics,   University of Oslo, N-0316 Oslo, Norway}
\altaffiltext{2}{Oslo College, Faculty of engineering, Pilestredet 35, N-0167 Oslo, Norway}

\maketitle

\begin{abstract}
We explore flat $\Lambda \textrm{CDM}$ models with bulk viscosity, and study the role of the bulk viscosity in the evolution of these universe models. The dynamical equations for these models are obtained and solved for some cases of bulk viscosity. We obtain differential equations for the Hubble parameter $H$ and the energy density of dark matter $\rho_{m}$, for which we give analytical solutions for some cases and for the general case we give a numerical solution. Also we calculate the statefinder parameters for this model and display them in the $s$-$r$-plane.
\end{abstract}


\section{Introduction}

A host of observations show that our Universe is currently undergoing an accelerating expansion. This has been confirmed by astronomical observations, such as observations of type Ia supernovae (SNIa) (see ~\cite{Riess} and ~\cite{Perlmutter}), observations of large scale structure (LSS) (see ~\cite{Tegmark} and ~\cite{Abazajian}) and measurements of the cosmic microwave background (CMB) anisotropy (see ~\cite{Spergel}, ~\cite{Spergel1}, ~\cite{Spergel2} and ~\cite{Bennett}).
Based on these observations, cosmologists, have accepted the idea of dark energy, which is a fluid with negative pressure making up around $70\%$ of the present universe energy content, to be responsible for this acceleration due to repulsive gravitation. There are many theoretical models of dark energy, but the Lorentz invariant vacuum energy (LIVE) which can be represented by a cosmological constant $\Lambda$, with a constant equation of state parameter $w = -1$, is still the preferred model. The so-called $\Lambda \textrm{CDM}$ model, which in a flat universe model contain both LIVE and cold dark matter (CDM),i.e. dust, is the simplest cosmological model that is in agreement with the observational data and can explain the accelerated expansion of the Universe. 

In hydrodynamics (see ~\cite{Vis} and ~\cite{Viscositiii}), viscosity is defined as a measure of the resistance to flow of a fluid, and is related to the velocity gradient. There are two types of viscosity, namely, shear and bulk viscosity. The shear viscosity characterizes a change in shape of a fixed volume of the fluid, whereas the bulk viscosity characterizes a change in volume of the fluid of a fixed shape. In particular, the bulk viscosity being therefore relevant only for compressible liquids. The shear viscosity represents the ability of particles to transport momentum. It is usual to use shear viscosity in connection with spacetime anisotropy, while isotropic cosmological models can have bulk viscosity.

Many cosmological models have perfect fluid behaviour because it is most simple to deal with this type of behaviour, and these models seem to be in good agreement with cosmological observations (see ~\cite{Riess} and ~\cite{Spergel}). But, on a more physical and realistic basis we can replace the energy-momentum tensor for the simplest perfect fluid by introducing cosmic viscosity. This modifies the equation of state of the cosmic fluid. A review of universe models with viscosity is given by ~\cite{brevik02}.

\section{Statefinder parameters}

The statefinder parameter pair $\left\{s,r\right\}$, was introduced by ~\cite{Sahni} and ~\cite{Alam}, and for a flat universe they are defined as

\begin{align}
r&\equiv \frac{\dddot{a}}{aH^{3}},\\
s&\equiv \frac{r-1}{3(q-1/2)}.\label{eq:hizz}
\end{align}

The statefinder diagnostic has a geometrical character, because it is constructed from the space-time metric directly, which is more universal than "physical" variables, which are model-dependent. Introducing the statefinder parameters is a natural next step beyond the Hubble parameter $H$ depending on $\dot{a}$ and the deceleration parameter $q$ depending on $\ddot{a}$. The definition of the deceleration parameter is 
\begin{equation}
q = -\frac{\ddot{a}}{aH^{2}}.
\end{equation}
Expressing the deceleration parameter and the statefinder parameters in terms of the Hubble parameter and its derivates with respect to cosmic time, we obtain (see reference ~\cite{Gron2005})
\begin{align}
q&=-1-\frac{\dot{H}}{H^{2}},\label{eq:hekke}\\
r&=1+3\frac{\dot{H}}{H^{2}}+\frac{\ddot{H}}{H^{3}},\label{eq:hekkeo}\\ 
s&=-\frac{2}{3H}\frac{3H\dot{H}+\ddot{H}}{3H^{2}+2\dot{H}}.\label{eq:hekkeohekkeo}
\end{align}

\section{Viscous Fluid}

The first theory of relativistic viscous fluid was presented by ~\cite{eckart}. As noted by ~\cite{singh} Eckart's theory deals with first order deviation from equilibrium, while neglected second order terms are necessary to prevent non-causal behavior. ~\cite{israel} and ~\cite{pavon} have developed a second order theory. ~\cite{brevik02} and ~\cite{maartens} have presented exhaustive reviews of research on cosmological models with non-causal and causal theories of viscous fluids, respectively. In particular Grøn gave a detailed discussion of Murphy's class of viscous cosmological models, ~\cite{murphy}. Bulk viscosity driven cosmic expansion with the Israel-Stewart theory have been investigated by ~\cite{zim1}, ~\cite{mak}, ~\cite{paul} and ~\cite{arab}. As noted by ~\cite{lepe}, although Eckart's theory presents some causality problems, it is the simplest alternative and has been widely considered in cosmology, as documented in Grøn's review. In the present work we shall extend some recent works based on the Eckart theory.

The energy momentum tensor with bulk viscosity is given by
\begin{equation}
T_{\mu \nu}=(\rho+p-\xi\theta)u_{\mu}u_{\nu}+(p-\xi\theta)g_{\mu\nu},\label{eq:en}
\end{equation}  
where $\xi$ is bulk viscosity, and $\theta$ is the expansion scalar, defined as $\theta\equiv 3H$. In order to discuss the dark energy properties, some authors (~\cite{WBXHU} and ~\cite{MGHUXHM}) have proposed a possible form of bulk viscosity as
\begin{equation}
\xi=\xi_{0}+\xi_{1}\frac{\dot{a}}{a}+\xi_{2}\frac{\ddot{a}}{a}, \label{eq:syv}
\end{equation}
where $\xi_{0}$, $\xi_{1}$ and $\xi_{2}$ are constants. 

The motivation for considering this bulk viscosity is that according to fluid mechanics the transportviscosity phenomenon is related to velocity and acceleration. Since the exact form of the viscosity is not known, we study cosmological consequences of a parameterized bulk viscosity which is a linear combination of three terms: a constant and two terms proportional to the 'velocity', $\dot{a}$, and the 'acceleration', $\ddot{a}$, respectively. Hence, in this paper we will investigate the $\Lambda \textrm{CDM}$ universe models with bulk viscosity given by eq. (\ref{eq:syv}).

\section{Viscous $\Lambda \textrm{CDM}$ models}

The $\Lambda\textrm{CDM}$ model consists of two components, the non-relativistic matter component, with subscript $m$, and the dark energy component which is assumed to be a Lorentz invariant vacuum energy, LIVE, given as the cosmological constant, $\Lambda$. The equation of state of the dark energy and the matter is given respectively by $p_{\Lambda}=-\rho_{\Lambda}$ and $p_{m}=0$. Thus, the total pressure and density are given by
\begin{align}
p&=p_{m}+p_{\Lambda}=-\rho_{\Lambda}, \label{eq:p} \\
\rho&=\rho_{m}+\rho_{\Lambda}.\label{eq:roh} 
\end{align}

Writing out the Einstein equations
\begin{equation}
R_{\mu \nu}-\frac{1}{2}Rg_{\mu \nu}=\kappa T_{\mu \nu}, \label{eq:to}
\end{equation} 
for the Friedmann-Robertson-Walker metric of a flat universe 
\begin{equation}
ds^{2}=-dt^{2}+a(t)^{2}\left[dx^{2}+dy^{2}+dz^{2}\right], \label{eq:tre}
\end{equation}
and using equation (\ref{eq:en}), we obtain
\begin{align}
\frac{\dot{a}^{2}}{a^{2}}&=\frac{\kappa}{3}\rho , \label{eq:fire}\\
\frac{\ddot{a}}{a}&=-\frac{\kappa}{6}\left(\rho+3p-9\xi H\right)\label{eq:fem}
\end{align}
In analogy to the density parameters of the dark matter and LIVE with present values
\begin{equation}
\Omega_{m0}=\frac{\kappa \rho_{m0}}{3H_{0}^{2}}, \qquad \Omega_{\Lambda 0}=\frac{\kappa \rho_{\Lambda 0}}{3H_{0}^{2}},
\end{equation}
we introduce a dimensionless $viscosity$ $parameter$, $\Omega_{\xi}$, with present value
\begin{equation}
\Omega_{\xi 0}=\frac{3\kappa \xi}{H_{0}}.
\end{equation}
From eqs.(\ref{eq:fire}) and (\ref{eq:fem}) then follow
\begin{equation}
\Omega_{\xi 0}=\Omega_{m0}-2\Omega_{\Lambda 0}-2q_{0}.
\end{equation}
This parameter is a measure of the importance of viscosity compared to the densities of matter and
energy. If there is no mechanism producing viscosity, $\Omega_{\xi 0}=0$ and $q_{0}=(1/2)(\Omega_{m0}-2\Omega_{\Lambda 0})$. Hence one may obtain information of the importance of viscosity in the cosmic fluid from measurements of $\Omega_{m0}$, $\Omega_{\Lambda 0}$ and $q_{0}$ as was shown by ~\cite{pavon2}. At the present time we have very accurate measurements of $\Omega_{m0}$ and $\Omega_{\Lambda 0}$, but there is a relatively great uncertainty in the kinematical measurements of the time variation of the Hubble parameter, i.e. of $q_{0}$. So at the present time we have no accurate information from such measurements about the importance of cosmic viscosity. ~\cite{zim3} have shown, however, that cosmic particle production can produce an effective bulk viscosity which may be of significance for the explanation of the accelerated expansion of the universe.

The continuity equation takes the form
\begin{equation}
\dot{\rho}=-3H(\rho+p-3\xi H), \label{eq:seks} 
\end{equation}

Inserting equation (\ref{eq:syv}) in the continuity equation we obtain
\begin{equation}
\dot{\rho}=-3H\left[\rho+p-3\left(\xi_{0}+\xi_{1}\frac{\dot{a}}{a}+\xi_{2}\frac{\ddot{a}}{a}\right) H\right]. \label{eq:aatte} 
\end{equation}
Using equations (\ref{eq:fire}) and (\ref{eq:fem}), we find the following differential equation for the Hubble parameter (the Raychaudhuri equation) 
\begin{equation}
\dot{H}\left(1-\frac{3}{2}\kappa\xi_{2}H\right)=\frac{3}{2}\left(\kappa\xi_{1}-1\right)H^{2}+\frac{3}{2}\kappa\xi_{2}
H^{3}+\frac{3}{2}\kappa\xi_{0}H+\frac{\kappa\rho_{\Lambda}}{2}.\label{eq:ni}
\end{equation}
In what follows we study $\Lambda\textrm{CDM}$ models with different cases of bulk viscosity.

\subsection{1. case: $\xi_{0}\neq 0$ and $\xi_{1}=\xi_{2}=0$}

\begin{figure}
  \centering
  \includegraphics[width=0.4\textwidth]{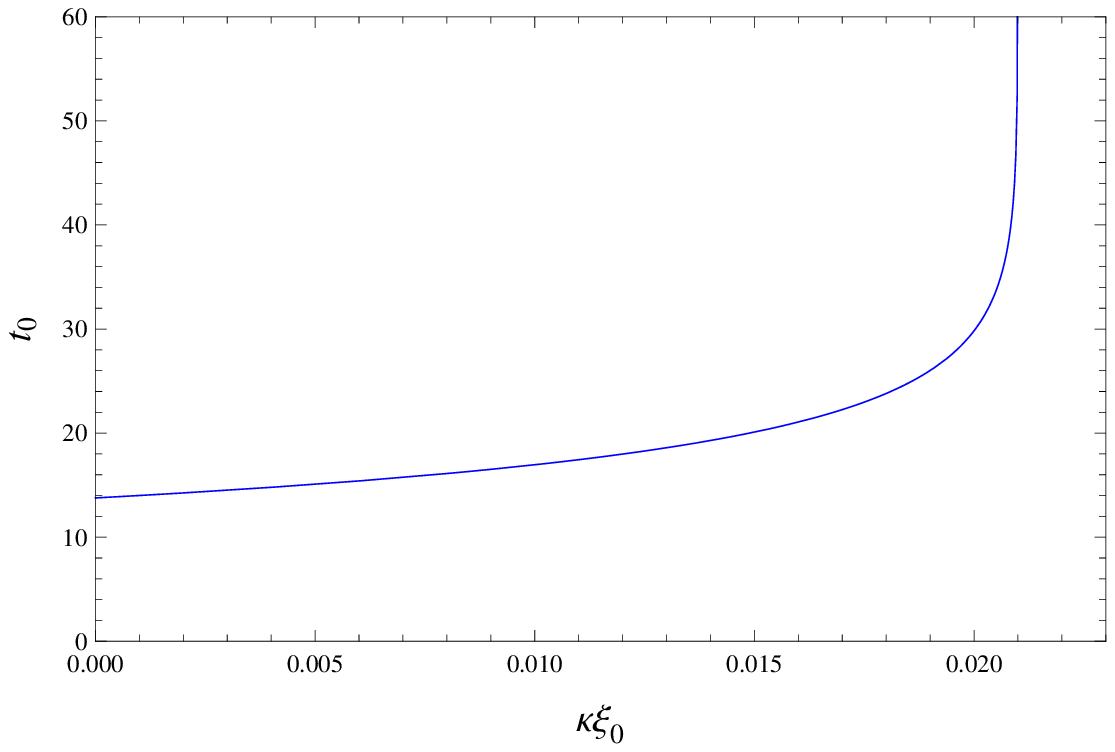}
  \caption[The $s$-$r$-plane for the $\Lambda \textrm{CDM}$ model with and without viscosity]{The age of the universe as a function of $\kappa\xi_{0}$. Here we have set $\Omega_{\Lambda 0}= 0.7$, $\Omega_{m 0}= 0.3$, and $H_{0} = 0.07$ which means that $t_{0}$ is given in billion years.}
  \label{fig:figtxi}
\end{figure}

In the present case the Raychaudhuri equation takes the form 
\begin{equation}
\dot{H}+\frac{3}{2}H^{2}-\frac{3}{2}\kappa\xi_{0} H-\frac{\kappa}{2}\rho_{\Lambda}=0. \label{eq:ti}
\end{equation}
The corresponding model with dark energy obeying the equation of state $p_{x}=w\rho_{x}$, $w\neq -1$, has been considered in ~\cite{brevik01} and ~\cite{gron1}.
Integrating equation (\ref{eq:ti}), we obtain
\begin{equation}
H(t)=\frac{\kappa\xi_{0}}{2}+\alpha\frac{1+C\textrm{e}^{-3\alpha(t-t_{0})}}{1-C\textrm{e}^{-3\alpha(t-t_{0})}},\label{eq:elve}
\end{equation}
where we have defined $\alpha$ and $C$ as
\begin{displaymath}
\alpha=\sqrt{\left(\frac{\kappa\xi_{0}}{2}\right)^{2}+H_{0}^{2}\Omega_{\Lambda 0}}, \qquad 
C=\frac{H_{0}-\frac{\kappa\xi_{0}}{2}-\alpha}{H_{0}-\frac{\kappa\xi_{0}}{2}+\alpha},
\end{displaymath}
where $H_{0}$ and $\Omega_{\Lambda 0}$ are the values of the Hubble parameter and the density parameter of the dark energy at the present time, $t_{0}$. Using that $H=\frac{\dot{a}}{a}$, we can rewrite equation (\ref{eq:elve}) as
\begin{equation}
\frac{da}{a}=\left[\frac{\kappa\xi_{0}}{2}+\alpha\frac{1+C\textrm{e}^{-3\alpha(t-t_{0})}}{1-C\textrm{e}^{-3\alpha(t-t_{0})}}\right]dt.\label{eq:tolv}
\end{equation}
Integrating this, we obtain
\begin{equation}
a(t)=a_{0}\textrm{e}^{\left(\frac{\kappa\xi_{0}}{2}+\alpha\right)(t-t_{0})}\left[\frac{1-C\textrm{e}^{-3\alpha(t-t_{0})}}{1-C}\right]^{2/3},\label{eq:tretten}
\end{equation}
where $a(t_{0})=a_{0}$, is the present scale factor. Demanding $a(0)=0$ and normalizing the scale factor to unity at the present time, we obtain
\begin{equation}
a(t)=\textrm{e}^{\frac{\kappa\xi_{0}}{2}(t-t_{0})}\left[\frac{\sinh\left(\frac{3}{2}\alpha t\right)}{\sinh\left(\frac{3}{2}\alpha t_{0}\right)}\right]^{\frac{2}{3}},\label{eq:trettena}
\end{equation}
which may be written
\begin{equation}
a(t)=A\textrm{e}^{\frac{\kappa\xi_{0}}{2}(t-t_{0})}\sinh^{\frac{2}{3}}{(\frac{3}{2}\alpha t)}, \qquad A=\left[\frac{H_{0}^{2}\Omega_{m0}-H_{0}\kappa\xi_{0}}{H_{0}^{2}\Omega_{\Lambda 0}+(\frac{\kappa\xi_{0}}{2})^{2}}\right]^{\frac{1}{3}}.
\end{equation}
The expression for the Hubble parameter, equation (\ref{eq:elve}), reduces to
\begin{equation}
H(t)=\frac{\kappa\xi_{0}}{2}+\alpha\coth{\left(\frac{3}{2}\alpha t\right)}.\label{eq:elvea}
\end{equation}
The age of the universe is given by 
\begin{equation}
t_{0}=(\frac{2}{3\alpha}) \operatorname{arcsinh}(1/A)^{3/2},\label{eq:tnull}
\end{equation}
which is plotted as a function of $\kappa\xi_{0}$ in Figure \ref{fig:figtxi} with the value of $H_{0}$ determined from the condition that $t_{0}=13,7\cdot 10^{9}$ years for $\xi_{0}=0$. It may be noted that the expression (\ref{eq:tnull}) is only valid for $\kappa \xi_{0}<H_{0}\Omega_{m 0}$. However realistic values of $\kappa\xi_{0}$ are much smaller than this (see ~\cite{brevikniten}). From this figure we see that when the viscosity increases, $t_{0}$ will also increase.  

\begin{figure}
  \centering
  \subfigure[][$a(t)$.]{\label{fig:fig2}\includegraphics[width=0.4\textwidth]{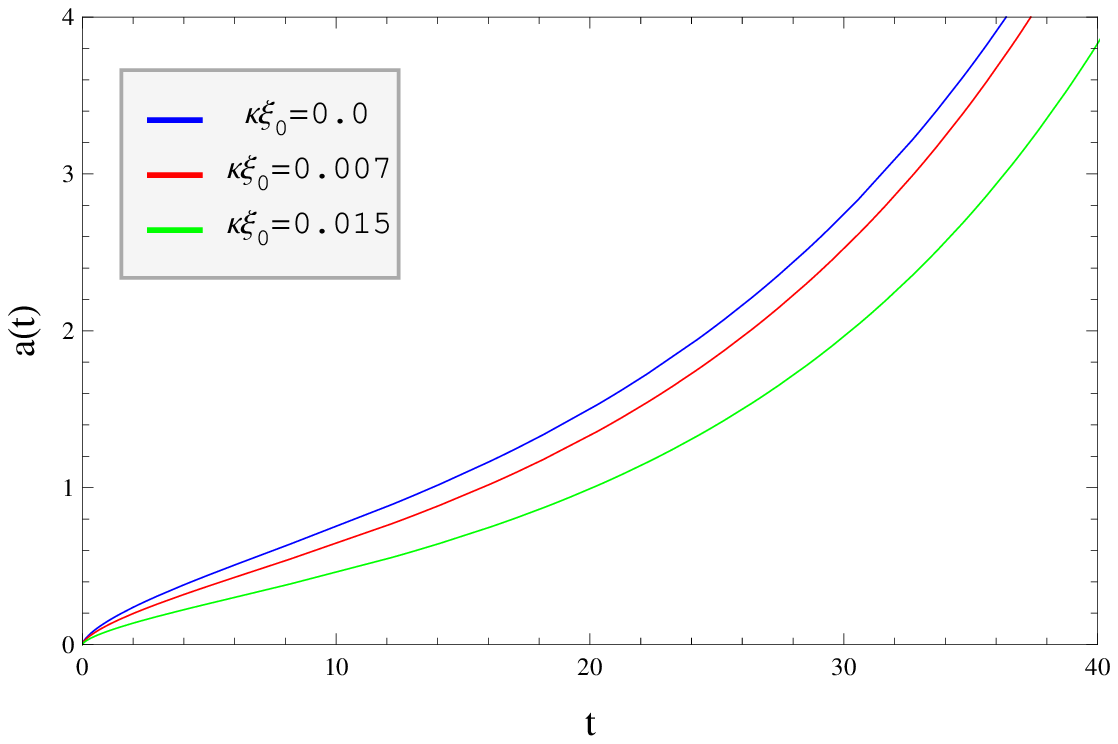}}                
  \subfigure[][$a(t)$.]{\label{fig:fig2a}\includegraphics[width=0.4\textwidth]{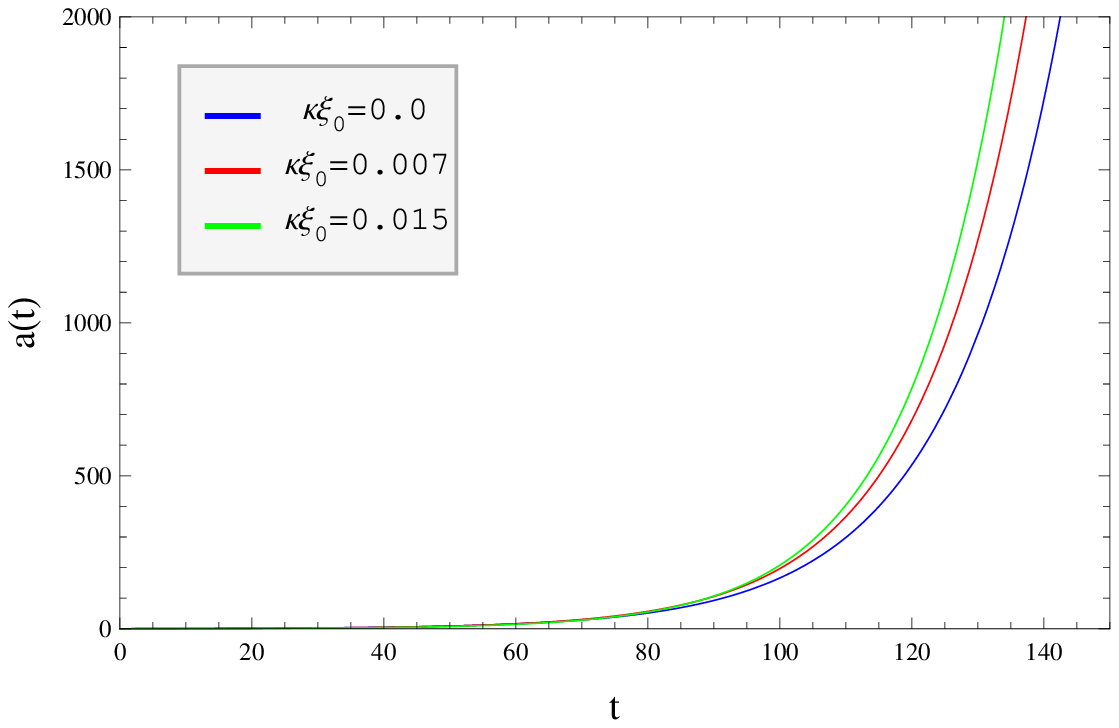}}
  \caption[The scale factor as a function of time for the $\Lambda \textrm{CDM}$ model with and without viscosity]{The scale factor as a function of time for the $\Lambda \textrm{CDM}$ model with and without viscosity. Here we have set $\Omega_{\Lambda 0}= 0.7$, $\Omega_{m 0}= 0.3$, and $H_{0} = 0.07$. The time is given in billion years.}
  \label{fig:fig1ca12}
\end{figure}

\begin{figure}
  \centering
  \subfigure[][$H(t)$.]{\label{fig:fig3}\includegraphics[width=0.4\textwidth]{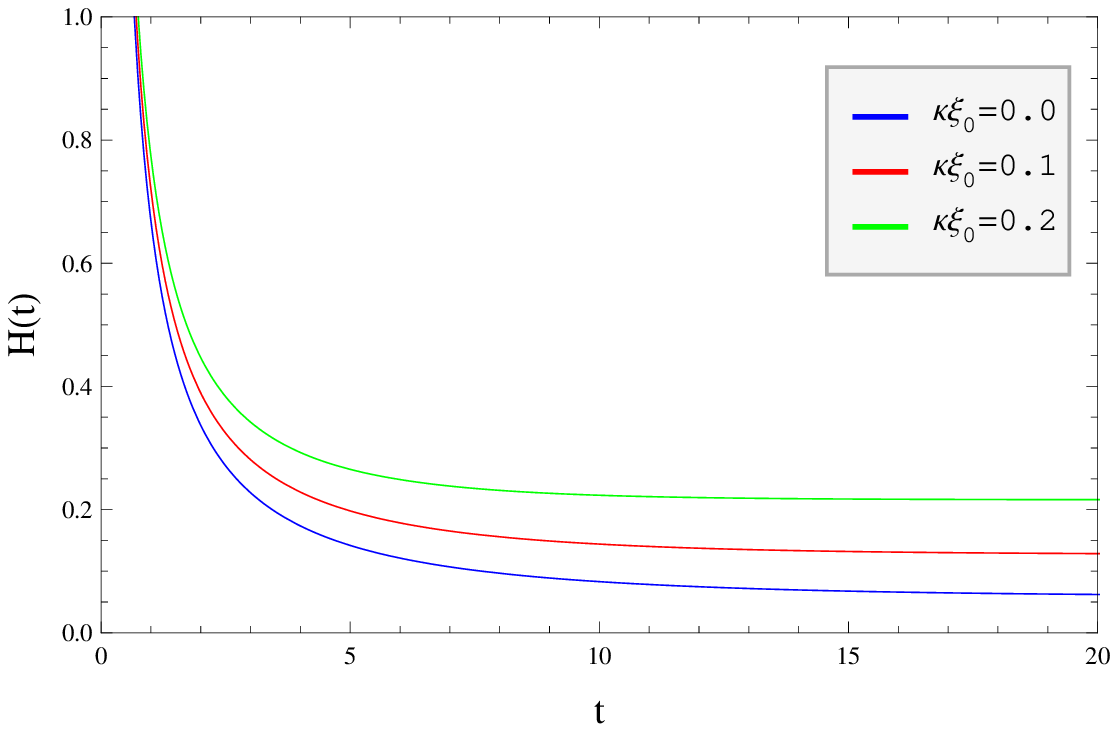}}
  \subfigure[][$\rho(t)$.]{\label{fig:fig4}\includegraphics[width=0.4\textwidth]{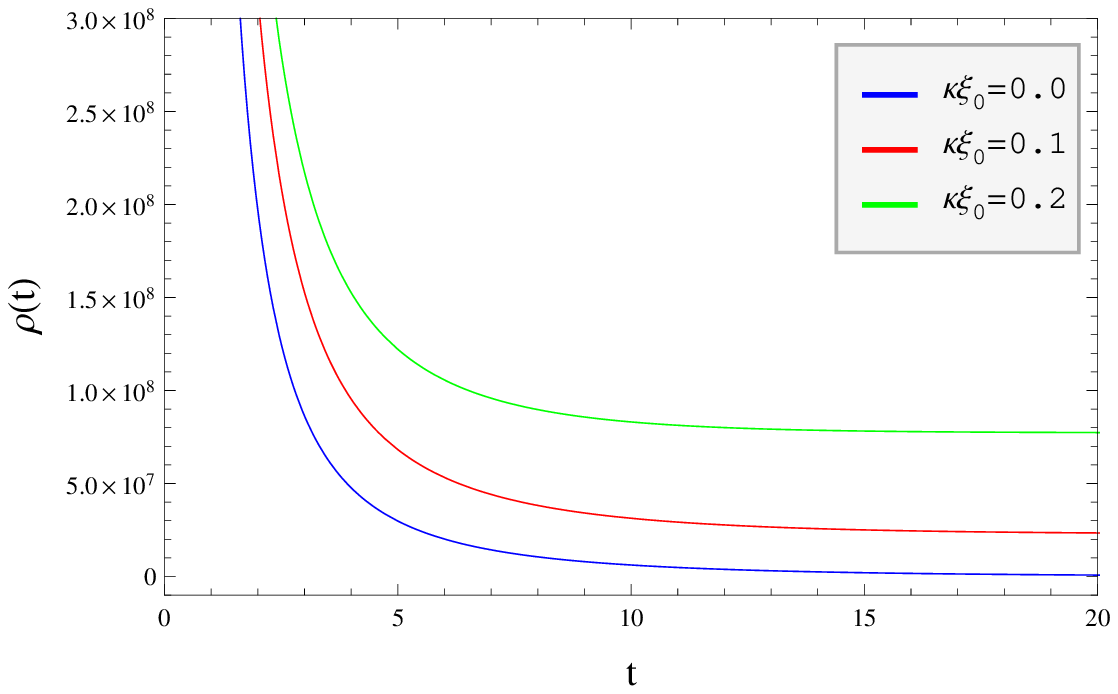}}
  \caption[The energy density for the non-relativistic matter as a function of time for the $\Lambda \textrm{CDM}$ model with and without viscosity]{The Hubble parameter and the energy density for the non-relativistic matter as a function of time for the $\Lambda \textrm{CDM}$ model with and without viscosity. Here we have set $\Omega_{\Lambda 0}= 0.7$, $\Omega_{m 0}= 0.3$, $H_{0} = 0.07$ and $c_{1}=1$. The time is given in billion years.}
  \label{fig:fig34}
\end{figure}

In the limit that the bulk viscosity goes to zero, we obtain (see ~\cite{gron12}) 
\begin{equation}
a(t)=\left(\frac{\Omega_{m0}}{\Omega_{\Lambda 0}}\right)^{\frac{1}{3}}\sinh^{\frac{2}{3}}\left(\frac{3}{2}\sqrt{\Omega_{\Lambda 0}}H_{0}t\right),\label{eq:trettenb}
\end{equation}
which is the expression for the scale factor for the $\Lambda \textrm{CDM}$ model with no viscosity. Here, $\Omega_{m0}$ is the present density parameter of matter. The continuity equation in this case has the form
\begin{equation}
\dot{\rho_{m}}=-3H\left(\rho_{m}-3\xi_{0}H\right).\label{eq:fjortten}
\end{equation}
Inserting eq.(\ref{eq:elvea}), and using the expression for $\alpha$ we find that the general solution of this equation is
\begin{equation}
\rho_{m}(t)=c_{1}\left[\frac{\kappa \xi_{0}}{\alpha}\cosh\left(3\alpha t\right)+2\sinh\left(3\alpha t\right)+c_{2}+c_{3}\textrm{e}^{\frac{-3}{2}\kappa \xi_{0}t}\right],
\end{equation}
where
\begin{equation}
c_{1}\equiv\frac{3\xi_{0}\alpha}{4\sinh^{2}\left(\frac{3}{2}\alpha t\right)}, \qquad c_{2}\equiv\frac{4H_{0}^{2}\Omega_{\Lambda 0}}{\kappa \xi_{0}\alpha}
\end{equation}
and $c_{3}$ is an integration constant. Note that 
\begin{equation}
\lim_{t\rightarrow\infty}\rho_{m}=3\xi_{0}\left(\alpha+\frac{\kappa\xi_{0}}{2}\right).
\end{equation}
Thus, due to the viscosity the density of the matter does not approach zero in the infinitely far future. In Figures \ref{fig:fig1ca12} and  \ref{fig:fig34} we have plotted the time dependent of the scale factor, $a(t)$, the Hubble parameter $H(t)$ and the energy density of matter $\rho(t)$ for different values of $\kappa\xi_{0}$. These figures show that the scale factor starts with zero and the Hubble parameter is infinitely large at beginning of the cosmic evolution, which shows that there is a point singularity at the initial epoch. Also the energy density diverges at this point. It means that the universe starts with a big bang. As $t$ increases the scale factor will increase exponentially, and, as $t\rightarrow \infty$, the scale factor becomes infinite, whereas the Hubble parameter and the energy density become finite. If the bulk viscosity is zero, the energy density tends to zero as $t\rightarrow \infty$. Therefore, this model will give an empty universe for large times $t$. For the $\Lambda\textrm{CDM}$ model with bulk viscosity, as $t\rightarrow \infty$ the energy density converges to a finite value. It means that for this model the energy density will stay constant for large times, $t$. The bigger the value of the bulk viscosity is, the bigger this constant value of the energy density will be.

Using the definitions in eqs.(\ref{eq:hekke})-(\ref{eq:hekkeohekkeo}) we find the following expressions for the deceleration parameter and the statefinder parameters
\begin{align}
q&=-1+\frac{3}{2}f^{2}(t),\\
r&=1-\frac{9}{4}\frac{\kappa \xi_{0}}{\alpha}f^{3}(t)\sinh{\left(\frac{3}{2}\alpha t\right)}\,\\
s&=\frac{\kappa \xi_{0}}{2\alpha}\frac{f^{3}(t)}{1-f^{2}(t)}\sinh{\left(\frac{3}{2}\alpha t\right)},
\end{align}
where we have defined
\begin{equation}
f(t)\equiv \frac{\alpha}{\frac{\kappa\xi_{0}}{2}\sinh{\left(\frac{3}{2}\alpha t\right)}+\alpha\cosh{\left(\frac{3}{2}\alpha t\right)}}.
\end{equation}

\begin{figure}
  \centering
  \subfigure[][$q(t)$.]{\label{fig:figq1}\includegraphics[width=0.4\textwidth]{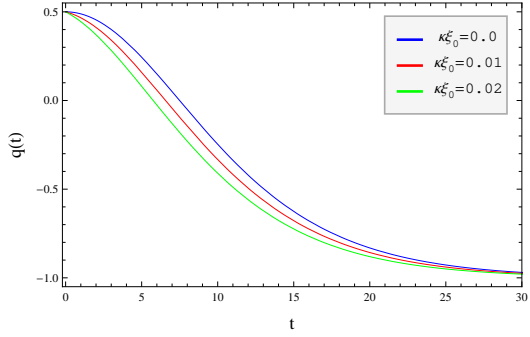}}
  \subfigure[][$s(t)$.]{\label{fig:figst}\includegraphics[width=0.4\textwidth]{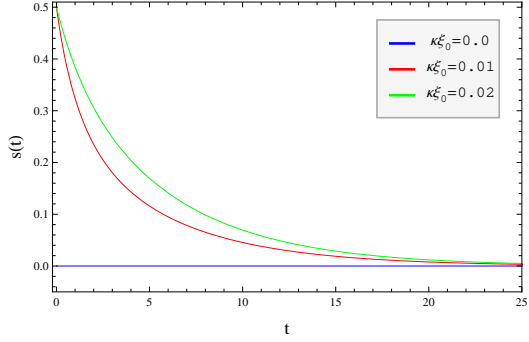}}
  \subfigure[][$r(t)$.]{\label{fig:figsr1}\includegraphics[width=0.4\textwidth]{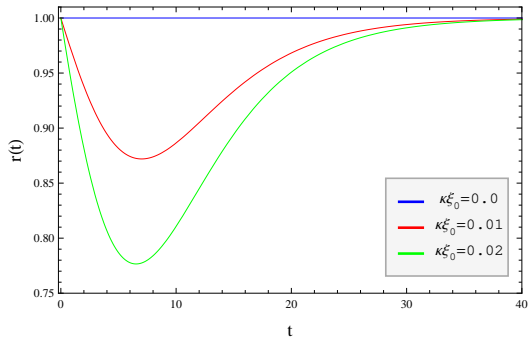}}
  \caption[The $s$-$r$-plane for the $\Lambda \textrm{CDM}$ model with and without viscosity]{The time evolution of the deceleration parameter and the statefinder parameters for the $\Lambda \textrm{CDM}$ model with and without viscosity have been displayed. Here we have set $\Omega_{\Lambda 0}= 0.7$, $\Omega_{m 0}= 0.3$, $H_{0} = 0.07$. The time is given in billion years.}
  \label{fig:figsr11}
\end{figure}

\begin{figure}
  \centering
  \includegraphics[width=0.4\textwidth]{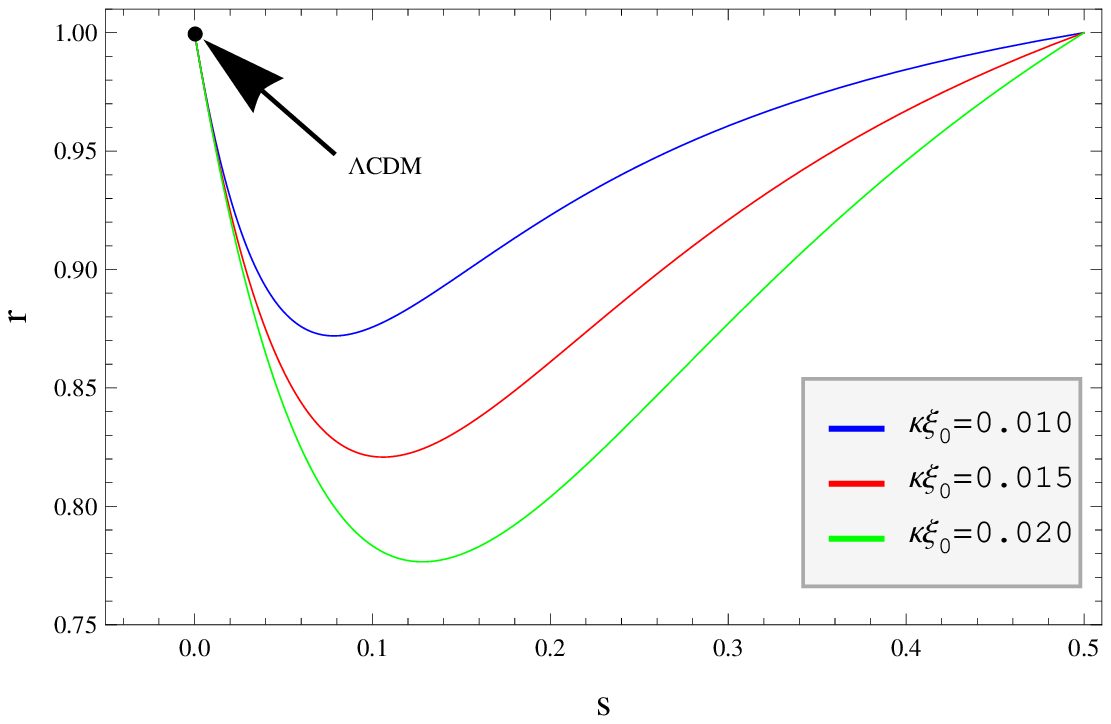}
  \caption[The $s$-$r$-plane for the $\Lambda \textrm{CDM}$ model with and without viscosity]{The $s$-$r$-plane for the $\Lambda \textrm{CDM}$ model with and without viscosity. The point $\{0,1\}$ corresponds to $\{s,r\}$ for the $\Lambda\textrm{CDM}$ model with no viscosity. Here we have set $\Omega_{\Lambda 0}= 0.7$, $\Omega_{m 0}= 0.3$, $H_{0} = 0.07$.}
  \label{fig:figsr12}
\end{figure}

We have plotted the time evolution of the deceleration parameter and the statefinder parameters in Fig.\ref{fig:figsr11}. The transition from deceleration to accelerated expansion happens at a point of time $t_{1}$, given by $q=0$, that is $f(t_{1})=\sqrt{2/3}$. From figure \ref{fig:figq1} we can see that the transition time to accelerated expansion is earlier the larger the viscosity is. From this figure we can see that the deceleration parameter starts at $q=0.5$ and as $t$ increases the deceleration parameter will decrease and as $t\rightarrow\infty$ it will approach its minimum value, which is $q=-1$. The bigger the value of the bulk viscosity is the faster the deceleration parameter will reach this value.

When the bulk viscosity is zero, i.e. $\xi_{0}=0$, the statefinder parameters have the values $\left\{s,r\right\}=\left\{0,1\right\}$, which is what we would expect for the $\Lambda \textrm{CDM}$ model. From these figures we can also see that for the $\Lambda \textrm{CDM}$ model with bulk viscosity, as $t\rightarrow\infty$ the statefinder parameters will approach $\left\{s,r\right\}=\left\{0,1\right\}$.

In Fig.\ref{fig:figsr12} we have plotted the $s$-$r$-plane for this model for different values of $\kappa\xi_{0}$. The point $\{0,1\}$ corresponds to $\{s,r\}$ for the $\Lambda\textrm{CDM}$ model with no viscosity.

\subsection{2. case: $\xi_{1}\neq 0$ and $\xi_{0}=\xi_{2}=0$}

In this case the Raychaudhuri equation takes the form
\begin{equation}
\dot{H}+\frac{3}{2}\left(1-\kappa\xi_{1}\right)H^{2}-\frac{\kappa}{2}\rho_{\Lambda}=0.\label{eq:femten}
\end{equation}
Assuming that $\kappa\xi_{1}<1$ and integrating this equation we obtain
\begin{equation}
H(t)=\hat{H}_{1}\frac{1+\hat{C}\textrm{e}^{-3\hat{H}_{2}(t-t_{0})}}{1-\hat{C}\textrm{e}^{-3\hat{H}_{2}(t-t_{0})}},\label{eq:seksten}
\end{equation}
where we have introduced
\begin{equation}
\hat{H}_{1}^{2}=\frac{H_{0}^{2}\Omega_{\Lambda 0}}{1-\kappa\xi_{1}}, \qquad  \hat{H}_{2}^{2}=H_{0}^{2}\Omega_{\Lambda 0}\left(1-\kappa\xi_{1}\right)\nonumber
\end{equation}
and
\begin{equation}
\hat{C}=\frac{H_{0}-\hat{H}_{1}}{H_{0}+\hat{H}_{1}} \nonumber.
\end{equation}
Using the definition of the Hubble parameter we can rewrite equation (\ref{eq:seksten}) as
\begin{equation}
\frac{da}{a}=\hat{H}_{1}\frac{1+\hat{C}\textrm{e}^{-3\hat{H}_{2}(t-t_{0})}}{1-\hat{C}\textrm{e}^{-3\hat{H}_{2}(t-t_{0})}}dt,\label{sytten}
\end{equation}
Integrating this equation and demanding that $a(0)=0$, we obtain
\begin{equation}
a(t)=\left[\frac{\sinh{\left(\frac{3}{2}\hat{H}_{2}t\right)}}{\sinh{\left(\frac{3}{2}\hat{H}_{2}t_{0}\right)}}\right]^{\frac{2}{3\left(1-\kappa\xi_{1}\right)}}.\label{eq:nitten}
\end{equation}
which may be written 
\begin{equation}
a(t)=A_{2}\sinh^{\frac{2}{3\left(1-\kappa\xi_{1}\right)}}{\left(\frac{3}{2}\hat{H}_{2}t\right)},
\end{equation}
where 
\begin{equation}
A_{2}=\left(\frac{1-\kappa\xi_{1}-\Omega_{\Lambda 0}}{\Omega_{\Lambda 0}}\right)^{\frac{1}{3\left(1-\kappa\xi_{1}\right)}}.\nonumber
\end{equation}
Normalizing the scale factor to unity at the present time, i.e., $a(t_{0})=1$, we find that the age of the universe, today, is given by
\begin{equation}
t_{0}=\frac{2}{3}\frac{1}{H_{0}\sqrt{\Omega_{\Lambda 0}\left(1-\kappa\xi_{1}\right)}}\textrm{arcsinh}\sqrt{\frac{\Omega_{\Lambda 0}}{1-\kappa\xi_{1}-\Omega_{\Lambda 0}}},
\end{equation}
which shows that the age of the universe increases with $\xi_{1}$ for a given value of $H_{0}$.

\begin{figure}
  \centering
  \includegraphics[width=0.4\textwidth]{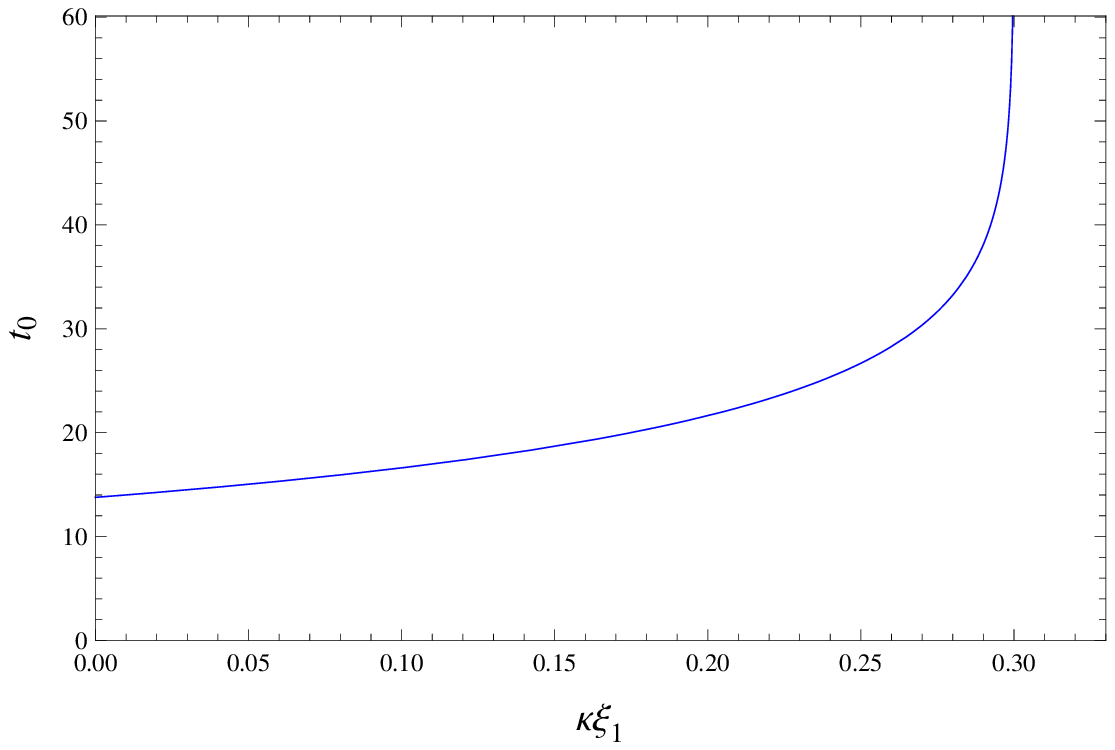}
  \caption[The $s$-$r$-plane for the $\Lambda \textrm{CDM}$ model with and without viscosity]{The age of the universe as a function of $\kappa\xi_{1}$. Here we have set $\Omega_{\Lambda 0}= 0.7$ and $H_{0} = 0.07$ which means that $t_{0}$ is given in billion years.}
  \label{fig:fig2ct}
\end{figure}

\begin{figure}
  \centering
  \subfigure[][$a(t)$.]{\label{fig:fig5}\includegraphics[width=0.4\textwidth]{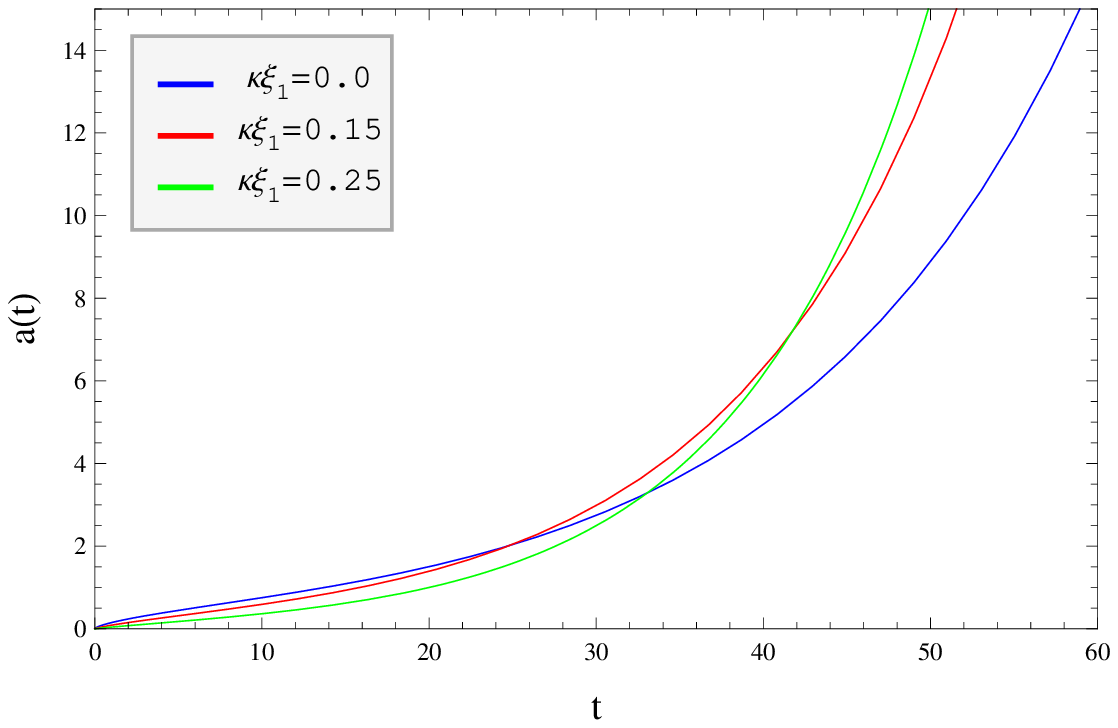}}          
  \subfigure[][$H(t)$.]{\label{fig:fig6}\includegraphics[width=0.4\textwidth]{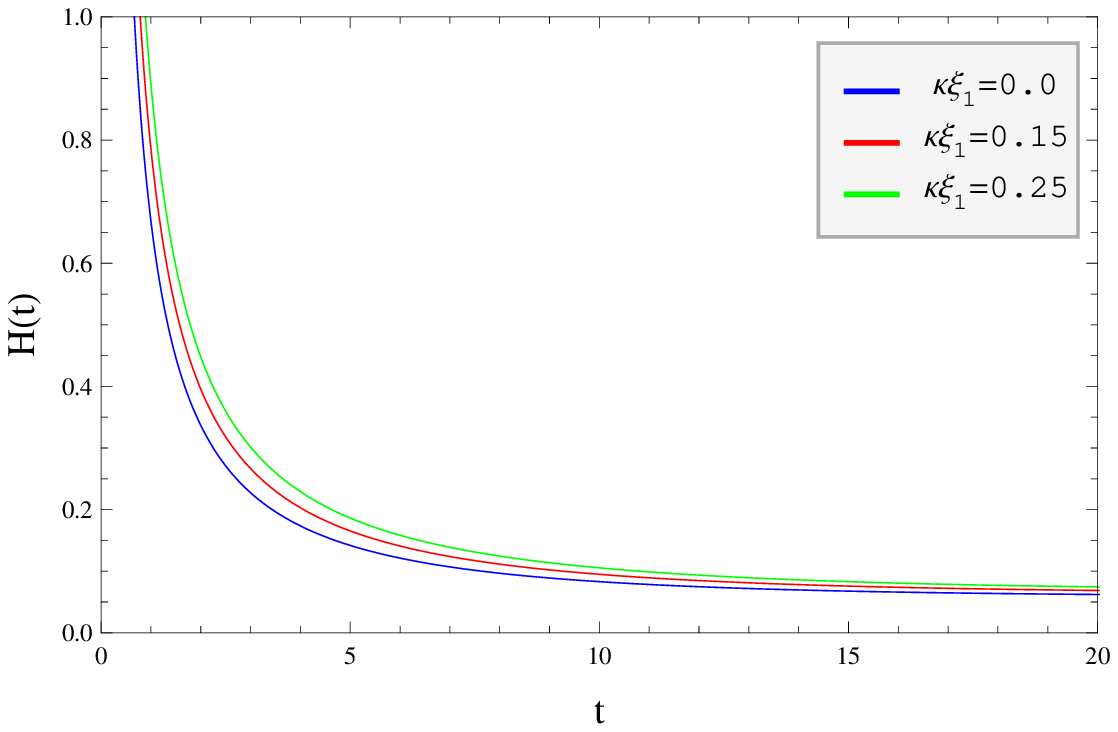}}
  \subfigure[][$\rho(t)$.]{\label{fig:fig7}\includegraphics[width=0.4\textwidth]{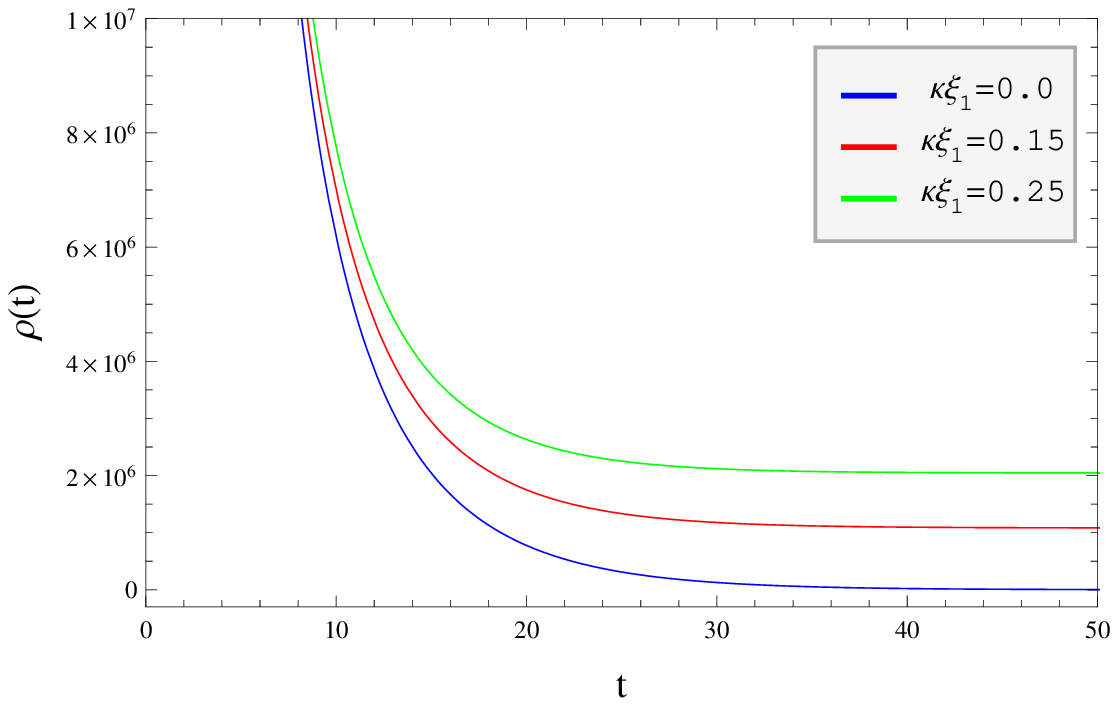}}
  \caption[The scale factor as a function of time for the $\Lambda \textrm{CDM}$ model with and without viscosity]{The scale factor, the Hubble parameter and the energy density of the non-relativistic matter as functions of time for the $\Lambda \textrm{CDM}$ model with and without viscosity are plotted in Fig.\ref{fig:fig5}, Fig.\ref{fig:fig6} and Fig. \ref{fig:fig7}, respectively. Here we have set $\Omega_{\Lambda 0}= 0.7$, $\Omega_{m 0}= 0.3$, $H_{0} = 0.07$. The time is given in billion years.}
  \label{fig:fig567}
\end{figure}

Using the expressions above we can reduce equation (\ref{eq:seksten}) to
\begin{equation}
H(t)=\hat{H}_{1}\coth{\left[\frac{3}{2}\hat{H}_{2}t\right]}.\label{eq:sekstento}
\end{equation}
In this case the continuity equation has the form
\begin{equation}
\dot{\rho_{m}}=-3H\left[(1-\kappa\xi_{1})\rho_{m}-\xi_{1}\kappa\rho_{\Lambda}\right].\label{eq:tjue}
\end{equation}
Inserting equation (\ref{eq:sekstento}) in equation (\ref{eq:tjue}) and integrating, gives the following expression for the energy density of non-relativistic matter
\begin{equation}
\rho_{m}(t)=\left(\rho_{m0}-\frac{3\xi_{1}H_{0}^{2}\Omega_{\Lambda 0}}{1-\kappa\xi_{1}}\right)\frac{1}{a(t)^{3\left(1-\kappa\xi_{1}\right)}}+\frac{3\xi_{1}H_{0}^{2}\Omega_{\Lambda 0}}{1-\kappa\xi_{1}},\label{eq:tjuen}
\end{equation}
where $\rho_{m0}$ is the present energy density of the matter. We also note here that 
\begin{equation}
\lim_{t\rightarrow\infty}\rho_{m}=\frac{3\xi_{1}H_{0}^{2}\Omega_{\Lambda 0}}{1-\kappa\xi_{1}}.
\end{equation}
It means that the density of the matter does not approach zero in the infinitely far future, due to viscosity. In Figure \ref{fig:fig567} we have plotted the time evolution of the scale factor $a(t)$, the Hubble parameter $H(t)$ and the energy density of the non-relativistic matter for different values of the bulk viscosity $\xi_{1}$. From these figures we can see that the universe starts with a big bang. Because, the scale factor starts with zero and the Hubble parameter and energy density are infinitely large at beginning of the cosmic evolution. As $t$ increases the scale factor will increase exponentially, and, as $t\rightarrow \infty$, the scale factor becomes infinite, whereas the Hubble parameter and the energy density become finite. If the bulk viscosity is zero, the energy density tends to zero as $t\rightarrow \infty$. Therefore, this model will also give an empty universe for large times $t$. For the $\Lambda\textrm{CDM}$ model with bulk viscosity, as $t\rightarrow \infty$ the energy density converges to a finite value. Just like in the first case, for this model the energy density will stay constant for large times, $t$, and the bigger the value of the bulk viscosity is, the bigger this constant value of the energy density will be.

The deceleration parameter and the statefinder parameters for this case take the form
\begin{align}
q&=-1+\frac{3}{2}\frac{\hat{H}_{2}}{\hat{H}_{1}}\frac{1}{\cosh^{2}{\left(\frac{3}{2}\hat{H}_{2}t\right)}},\\
r&=1-\frac{9}{2}\frac{\kappa\xi_{1}\left(1-\kappa\xi_{1}\right)}{\cosh^{2}{\left(\frac{3}{2}\hat{H}_{2}t\right)}},\\
s&=\frac{\kappa\xi_{1}\left(1-\kappa\xi_{1}\right)}{\kappa\xi_{1}+\sinh^{2}{\left(\frac{3}{2}\hat{H}_{2}t\right)}}.
\end{align}

\begin{figure}
  \centering
  \subfigure[][$q(t)$.]{\label{fig:figq2}\includegraphics[width=0.4\textwidth]{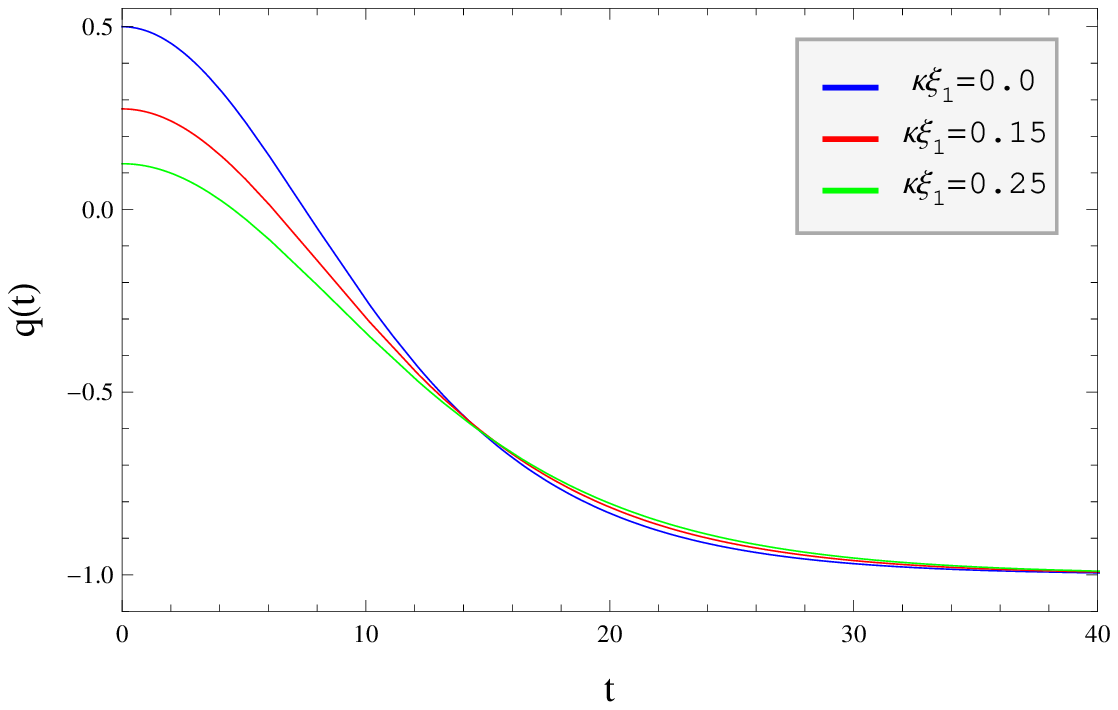}}        
  \subfigure[][$r(t)$.]{\label{fig:figr2}\includegraphics[width=0.4\textwidth]{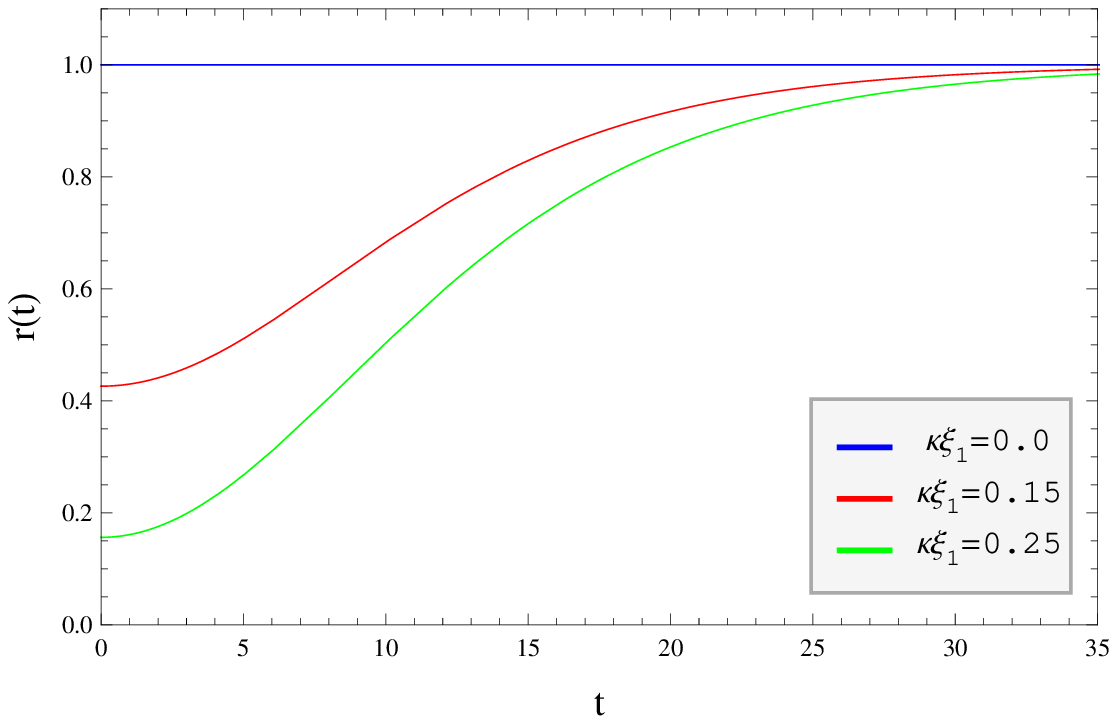}}
  \subfigure[][$s(t)$.]{\label{fig:figs2}\includegraphics[width=0.4\textwidth]{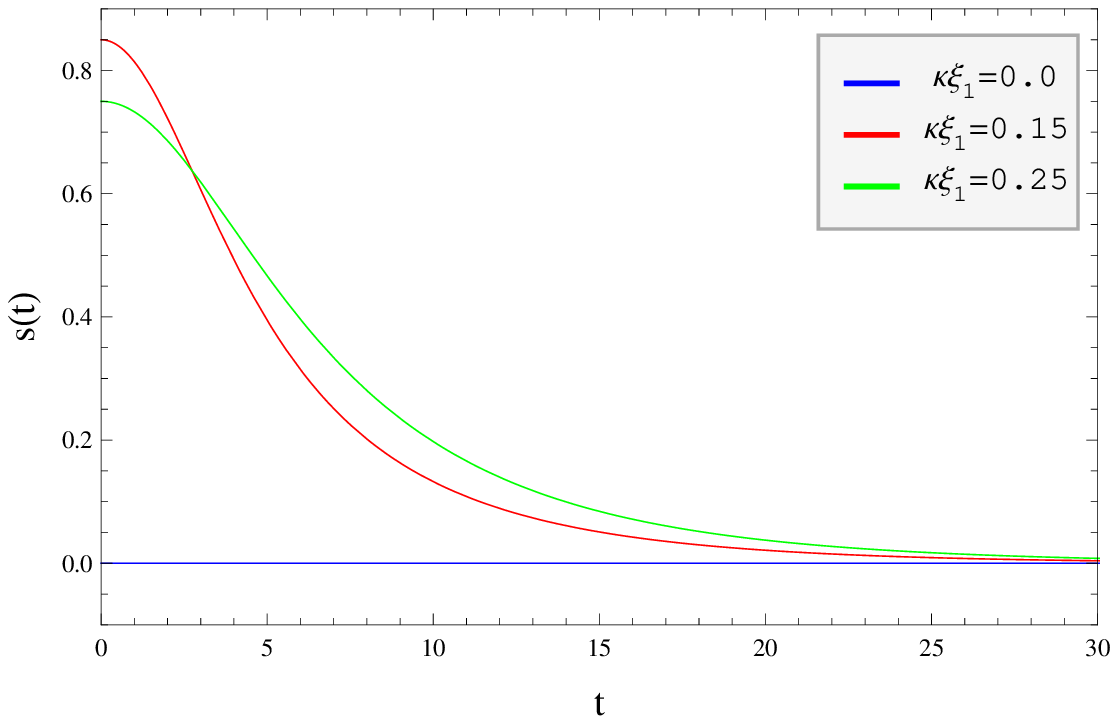}}
  \caption[The scale factor as a function of time for the $\Lambda \textrm{CDM}$ model with and without viscosity]{The deceleration parameter and the statefinder parameters $r$ and $s$ as functions of time for the $\Lambda \textrm{CDM}$ model with and without viscosity in Fig.\ref{fig:figr2}, Fig.\ref{fig:figs2} and Fig. \ref{fig:figq2}, respectively. Here we have set $\Omega_{\Lambda 0}= 0.7$, $\Omega_{m 0}= 0.3$, $H_{0} = 0.07$. The time is given in billion years.}
  \label{fig:fig5678}
\end{figure}

\begin{figure}
  \centering
  \includegraphics[width=0.4\textwidth]{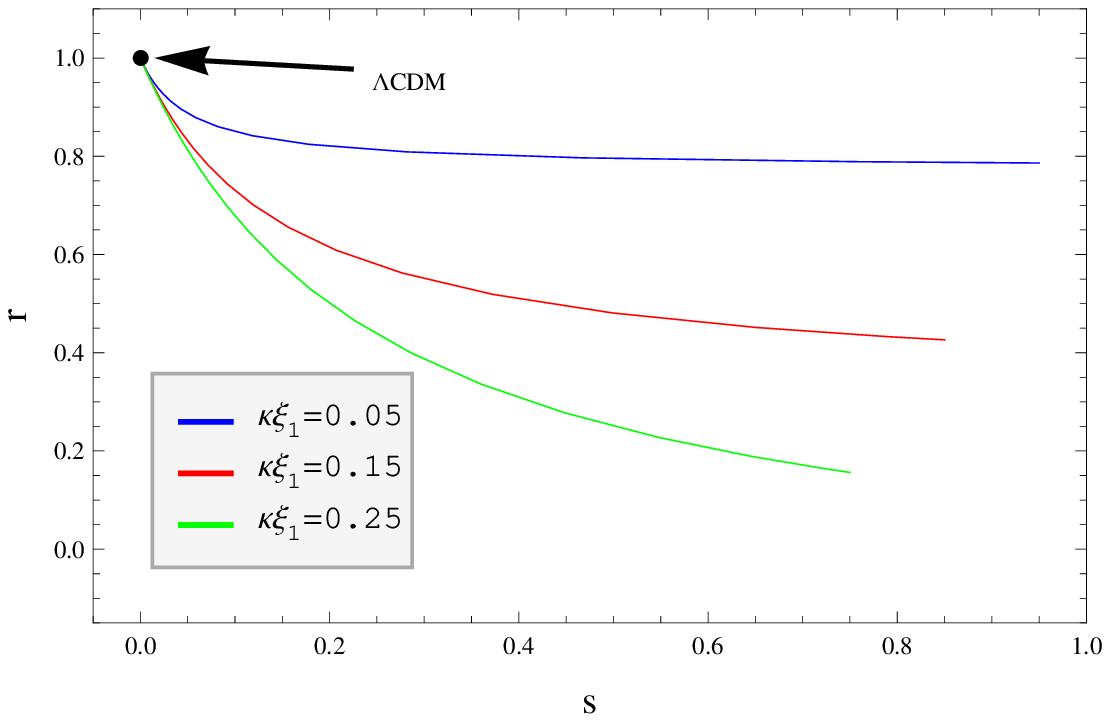}
  \caption[The $s$-$r$-plane for the $\Lambda \textrm{CDM}$ model with and without viscosity]{The $s$-$r$-plane for the $\Lambda \textrm{CDM}$ model with and without viscosity has been displayed. The point $\{0,1\}$ corresponds to $\{s,r\}$ for the $\Lambda\textrm{CDM}$ model with no viscosity.Here we have set $\Omega_{\Lambda 0}= 0.7$, $\Omega_{m 0}= 0.3$, $H_{0} = 0.07$.}
  \label{fig:figsr123}
\end{figure}

We have plotted the time evolution of the deceleration parameter and the statefinder parameters in Fig.\ref{fig:fig5678}. From these figures we can see that as $t$ increases the deceleration parameter will decrease, and as $t\rightarrow\infty$ it will approach its minimum value, which is $q=-1$. The bigger the value of the bulk viscosity is the faster the deceleration parameter will reach this value.

From Figure \ref{fig:figr2} and Figure \ref{fig:figs2} we can see that the statefinder parameter $r$ starts with a negative value but $s$ starts with a positive value, and as $t$ increases $r$ will increase but $s$ will decrease. We also see that as $t\rightarrow\infty$ the statefinder parameters will approach $\left\{s,r\right\}=\left\{0,1\right\}$. When the bulk viscosity is zero, i.e. $\xi_{1}=0$, the statefinder parameters have the values $\left\{s,r\right\}=\left\{0,1\right\}$, which is what we would expect for the $\Lambda \textrm{CDM}$ model with no viscosity.  

In Fig.\ref{fig:figsr123} we have plotted the $s$-$r$-plane for this model for different values of the bulk viscosity $\xi_{1}$. From this figure we can see that the graphs in the $s$-$r$-plane are parabola-like, and the statefinder parameters will eventually approach the point $\left\{s,r\right\}=\left\{0.1\right\}$. The point $\{0,1\}$ corresponds to $\{s,r\}$ for the $\Lambda\textrm{CDM}$ model with no viscosity.

\subsection{3. case: $\xi_{0}\neq 0$, $\xi_{1}\neq 0$ and $\xi_{2}=0$}

The Raychaudhuri equation in this case takes the form 
\begin{equation}
\dot{H}+\frac{3}{2}\left(1-\kappa\xi_{1}\right)H^{2}-\frac{3}{2}\kappa\xi_{0}H-\frac{\kappa}{2}\rho_{\Lambda}=0.\label{eq:tjuto}
\end{equation}
We assume $\kappa\xi_{1}<1$, and introduce
\begin{displaymath}
A=-(1-\kappa\xi_{1}), \qquad B=\kappa\xi_{0} \qquad \textrm{and} \qquad C=H_{0}^{2}\Omega_{\Lambda 0}.
\end{displaymath}
Integrating equation (\ref{eq:tjuto}) we obtain the following solution
\begin{equation} 
H(t)=-\frac{B}{2A}-\frac{\hat{\alpha}}{2A}\tanh{\left[\frac{3}{4}\hat{\alpha}(t-t_{0})-\phi_{0}\right]}, \label{tjutre}
\end{equation}
where we have defined
\begin{displaymath}
\hat{\alpha}=\sqrt{B^{2}-4AC}, \qquad \phi_{0}=\operatorname{arctanh}\left[\frac{2AH_{0}+B}{\hat{\alpha}}\right].   
\end{displaymath}
The corresponding expression for the scale factor has the form
\begin{equation}
a(t)=\textrm{e}^{-\frac{B}{2A}(t-t_{0})}\left(\frac{\cosh{\left[\frac{3}{4}\hat{\alpha}(t-t_{0})-\phi_{0}\right]}}{\cosh{\left[\phi_{0}\right]}}\right)^{-2/3A},\label{tjuseks}
\end{equation}
We find the following expressions for the deceleration parameter and the statefinder parameters
\begin{align}
q&=-1+\frac{3}{2}Ag^{2}(t),\\
r&=1+\frac{9}{2}Ag^{2}(t)\left[Ag(t)\sinh\left[\frac{3}{4}\hat{\alpha}(t-t_{0})-\phi_{0}\right]-1\right],\\
s&=\frac{Ag^{2}(t)\left[Ag(t)\sinh\left[\frac{3}{4}\hat{\alpha}(t-t_{0})-\phi_{0}\right]-1\right]}{Ag^{2}(t)-1},
\end{align}
where we have defined
\begin{equation}
g(t)\equiv \frac{\hat{\alpha}}{\hat{\alpha}\sinh{\left[\frac{3}{4}\hat{\alpha}(t-t_{0})-\phi_{0}\right]}-B\cosh{\left[\frac{3}{4}\hat{\alpha}(t-t_{0})-\phi_{0}\right]}}.
\end{equation}
The continuity equation for this case takes the form
\begin{equation}
\dot{\rho}_{m}=-3H\left(\rho_{m}-3\xi_{0}H-3\xi_{1}H^{2}\right).\label{eq:tjuni}
\end{equation}
We will only give a numerical solution to this equation.

\begin{figure}
  \centering
  \subfigure[][$a(t)$.]{\label{fig:fig8}\includegraphics[width=0.4\textwidth]{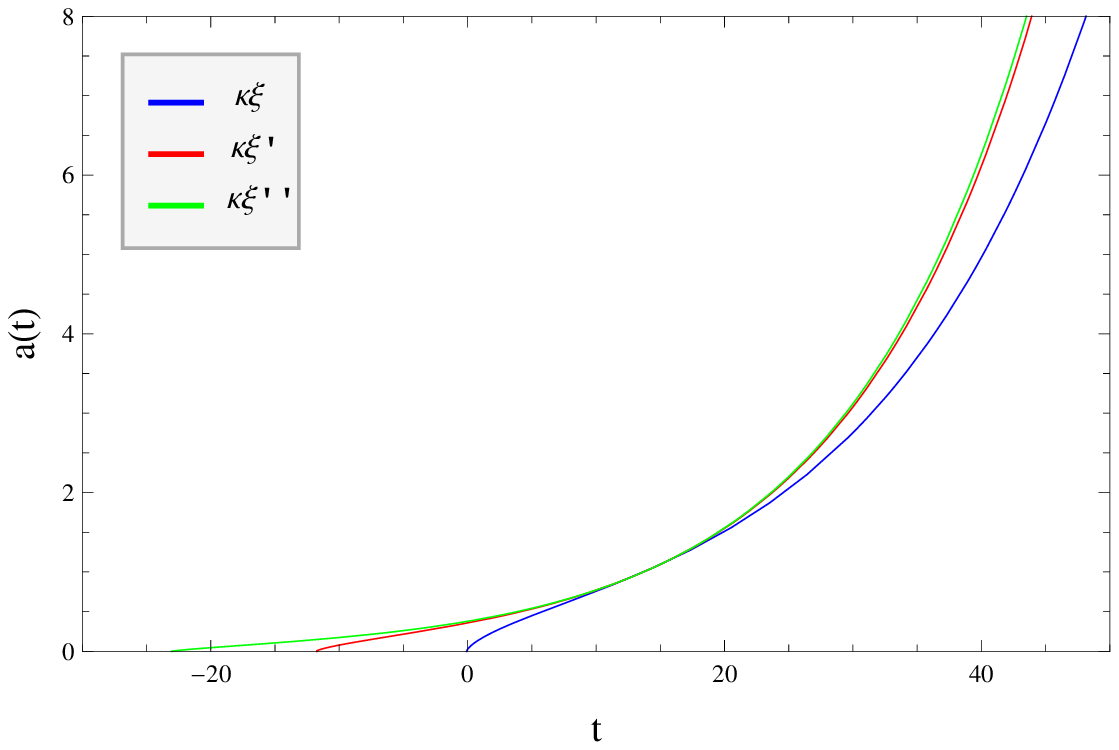}}          
  \subfigure[][$H(t)$.]{\label{fig:fig9}\includegraphics[width=0.4\textwidth]{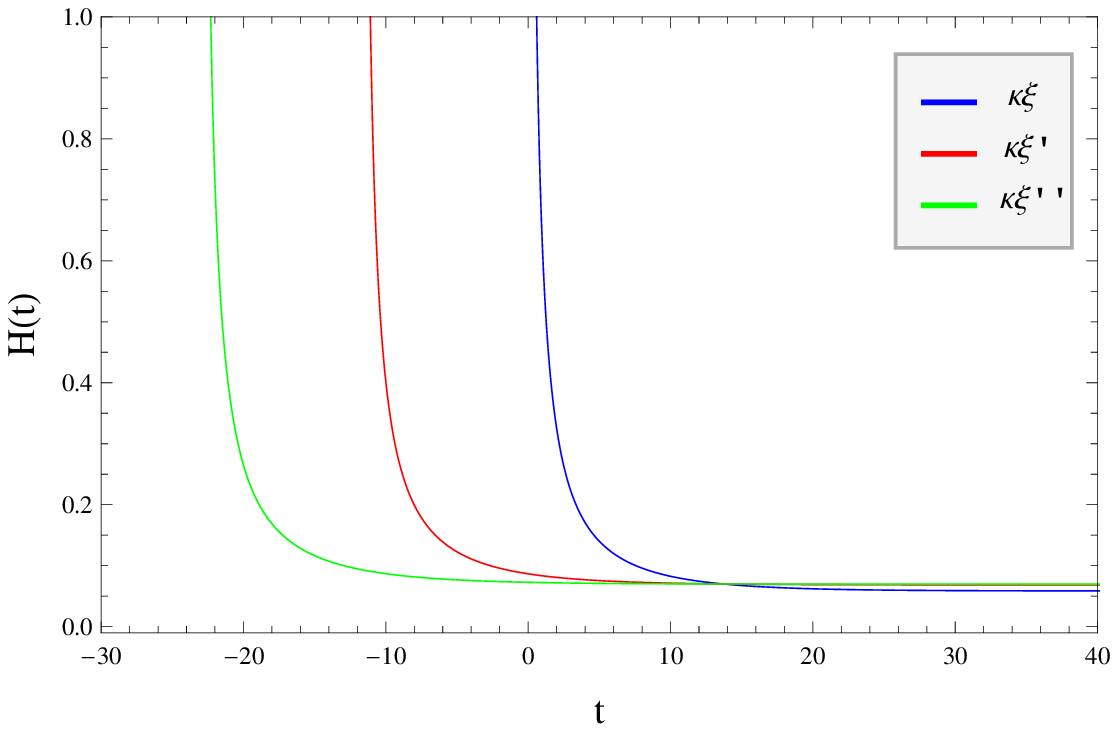}}          
  \subfigure[][$\rho(t)$.]{\label{fig:fig10}\includegraphics[width=0.4\textwidth]{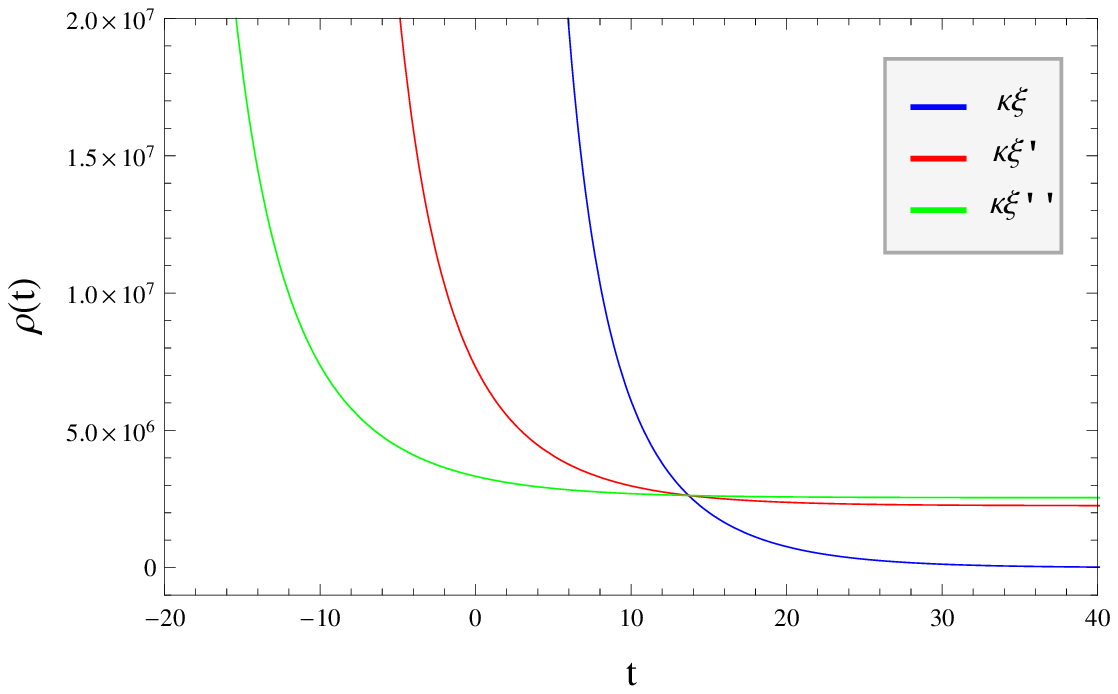}}
  \caption[The scale factor as a function of time for the $\Lambda \textrm{CDM}$ model with and without viscosity]{The time evolution of the scale factor, the Hubble parameter and the energy density for the non-relativistic matter for the $\Lambda \textrm{CDM}$ model with and without viscosity is plotted in Fig.\ref{fig:fig8}, \ref{fig:fig9} and \ref{fig:fig10}, respectively. Here $\kappa\xi:$ $\kappa\xi_{0}=.0$, $\kappa\xi_{1}=.0$, $\kappa\xi^{'}:$ $\kappa\xi_{0}=.015$, $\kappa\xi_{1}=.05$ and $\kappa\xi^{''}:$ $\kappa\xi_{0}=.01$, $\kappa\xi_{1}=.15$. Here we have set $\Omega_{\Lambda 0}= 0.7$, $\Omega_{m 0}= 0.3$, $H_{0} = 0.07$. The time is given in billion years.}
  \label{fig:fig8910}
\end{figure}

The time evolution for the scale factor, $a(t)$, the Hubble parameter, $H(t)$, and the energy density of the non-relativistic matter, $\rho(t)$ have been plotted in Figure \ref{fig:fig8910}. From Figure \ref{fig:fig9} and Figure \ref{fig:fig10} we see that for models with viscosity the Hubble parameter and the energy density diverge at a point of time $t<0$ and the bigger the viscosity is the earlier this point of time will be. The scale factor is zero at this point. Therefore the universe starts with a big bang. As $t$ increases the scale factor will increase exponentially, and the bigger the viscosity is the faster the universe will expand. The energy density will decrease as $t$ increases and it will approach a finite value as $t \rightarrow \infty$. The bigger the viscosity is the bigger this finite value will be.

\begin{figure}
  \centering
  \subfigure[][$q(t)$.]{\label{fig:figc3q}\includegraphics[width=0.4\textwidth]{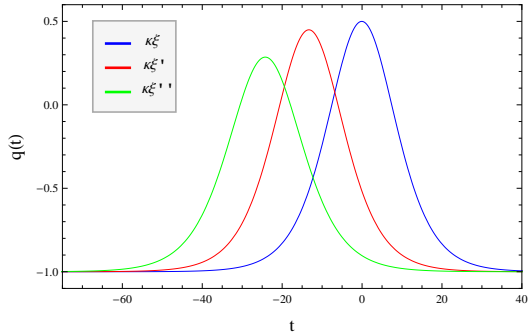}}        
  \subfigure[][$r(t)$.]{\label{fig:figc3r}\includegraphics[width=0.4\textwidth]{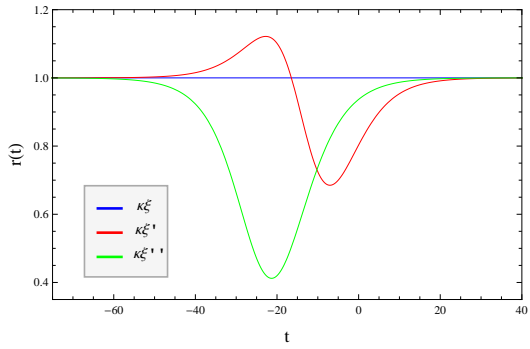}}
  \subfigure[][$s(t)$.]{\label{fig:figc3s}\includegraphics[width=0.4\textwidth]{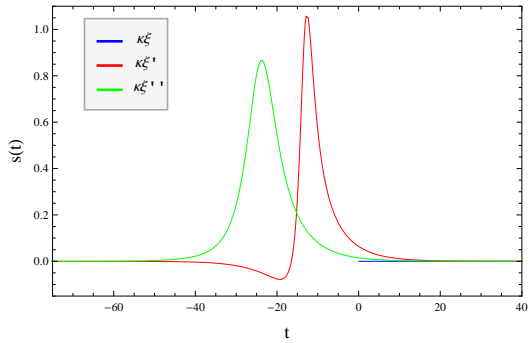}}
  \caption[The scale factor as a function of time for the $\Lambda \textrm{CDM}$ model with and without viscosity]{The time evolution of the deceleration parameter and the statefinder parameters for the $\Lambda \textrm{CDM}$ model with and without viscosity is plotted in Fig.\ref{fig:figc3q}, \ref{fig:figc3r} and \ref{fig:figc3s}, respectively. Here $\kappa\xi:$ $\kappa\xi_{0}=.0$, $\kappa\xi_{1}=.0$, $\kappa\xi^{'}:$ $\kappa\xi_{0}=.015$, $\kappa\xi_{1}=.05$ and $\kappa\xi^{''}:$ $\kappa\xi_{0}=.01$, $\kappa\xi_{1}=.15$. Here we have set $\Omega_{\Lambda 0}= 0.7$, $\Omega_{m 0}= 0.3$, $H_{0} = 0.07$. The time is given in billion years.}
  \label{fig:figc3qrs}
\end{figure}

The time evolution of the deceleration parameter and the statefinder parameters are plotted in Figure \ref{fig:figc3qrs}. From these figures we see that for these universe models as $t\rightarrow -\infty$ and as $t\rightarrow\infty$ the statefinder parameters will approach the same values as those of the $\Lambda \textrm{CDM}$ model with no viscosity, namely, $\left\{s,r\right\}=\left\{0,1\right\}$. 

\begin{figure}
  \centering
  \includegraphics[width=0.4\textwidth]{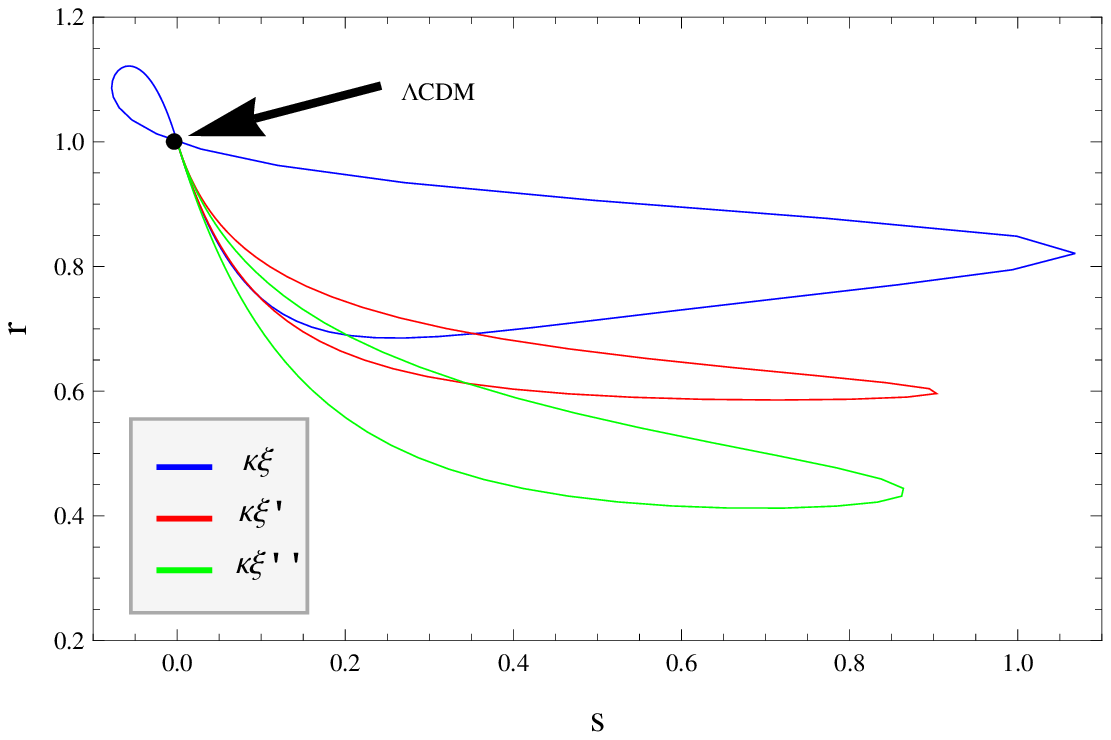}
  \caption[The $s$-$r$-plane for the $\Lambda \textrm{CDM}$ model with and without viscosity]{The $s$-$r$-plane for the $\Lambda \textrm{CDM}$ model with and without viscosity. Here $\kappa\xi:$ $\kappa\xi_{0}=.015$, $\kappa\xi_{1}=.05$, $\kappa\xi^{'}:$ $\kappa\xi_{0}=.005$, $\kappa\xi_{1}=.1$ and $\kappa\xi^{''}:$ $\kappa\xi_{0}=.01$, $\kappa\xi_{1}=.15$. Here we have set $\Omega_{\Lambda 0}= 0.7$, $\Omega_{m 0}= 0.3$, $H_{0} = 0.07$.}
  \label{fig:figc3rs}
\end{figure}

In Fig.\ref{fig:figc3rs} we have plotted $r$ as a function of $s$ for different values of $\xi_{0}$ and $\xi_{1}$. This figure shows that for small values of $\xi_{0}$ and $\xi_{1}$, when $t<0$, the graphs in the $s$-$r$-plane start from the point $\left\{0,1\right\}$ and as $t$ increases they  will go away from this point. For $t>0$ and as $t$ increases the graphs go toward the point $\left\{0,1\right\}$ and they will eventually approach this point. For bigger values of $\xi_{0}$ and $\xi_{1}$, we see that the graphs in the $s$-$r$-plane make two connected and closed loops. These graphs will at some point of time go through the point $\{0,1\}$.

\subsection{4. case: $\xi_{0}\neq 0$, $\xi_{1}\neq 0$ and $\xi_{2}\neq 0$}

\begin{figure}
  \centering
  \subfigure[][$a(t)$.]{\label{fig:figc4a}\includegraphics[width=0.4\textwidth]{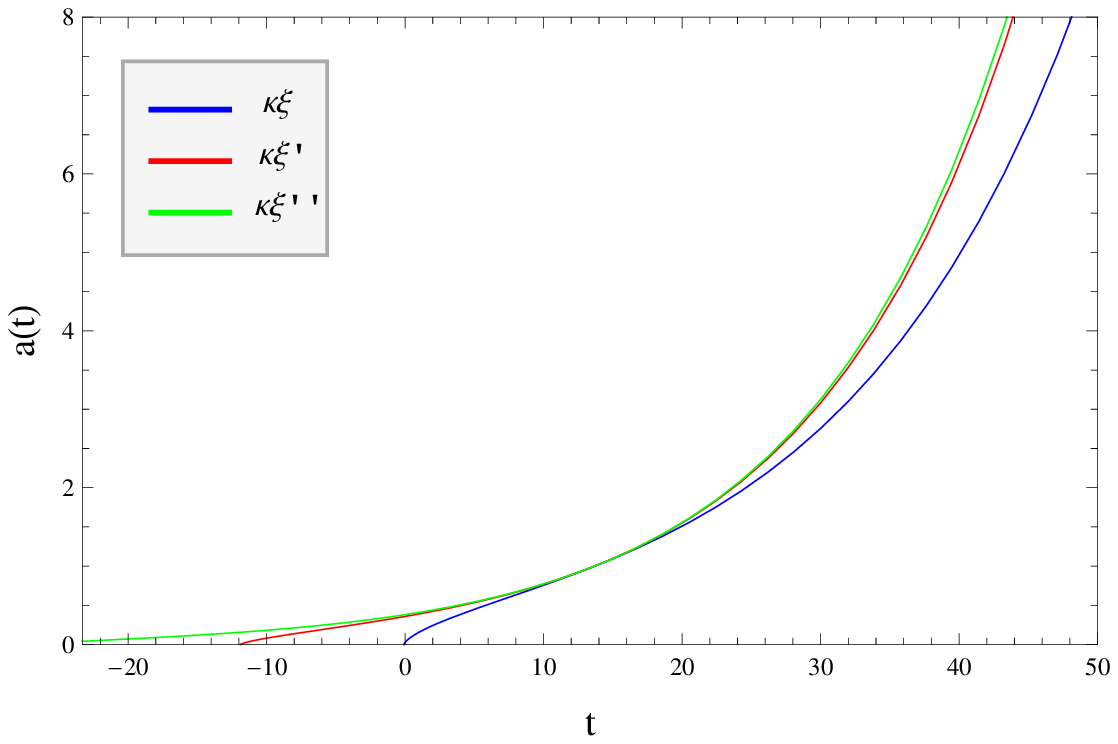}} 
  \subfigure[][$H(t)$.]{\label{fig:fighgenn}\includegraphics[width=0.4\textwidth]{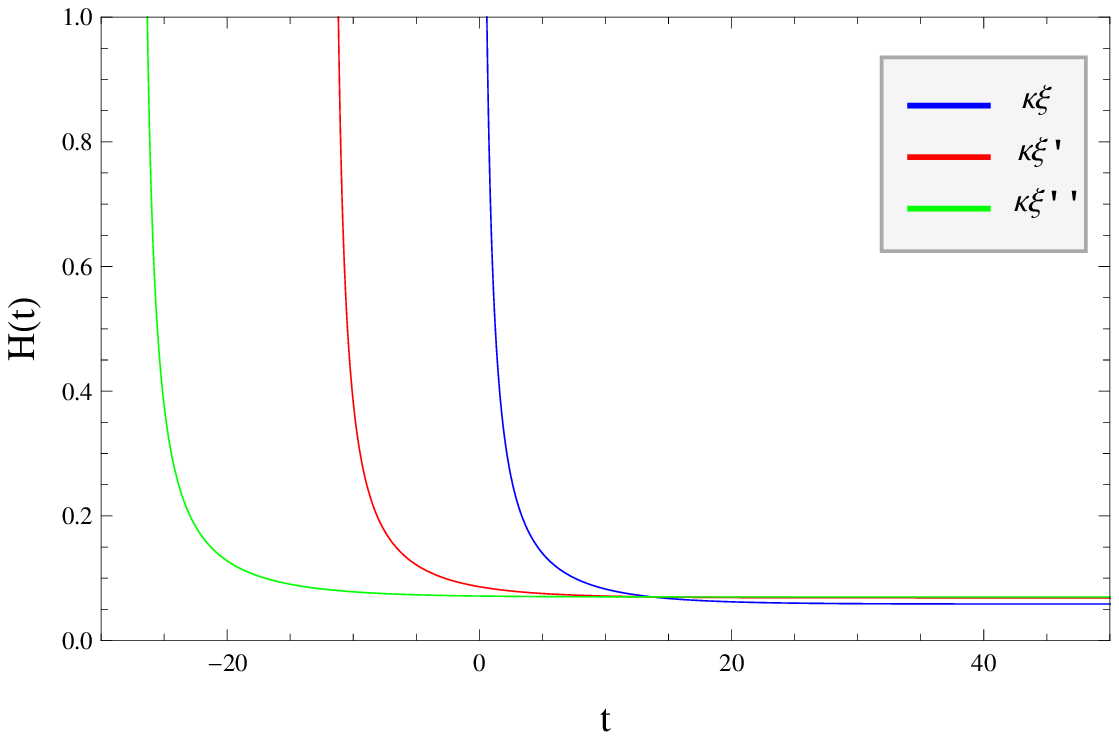}}      \subfigure[][$\rho(t)$.]{\label{fig:figrhogenn}\includegraphics[width=0.4\textwidth]{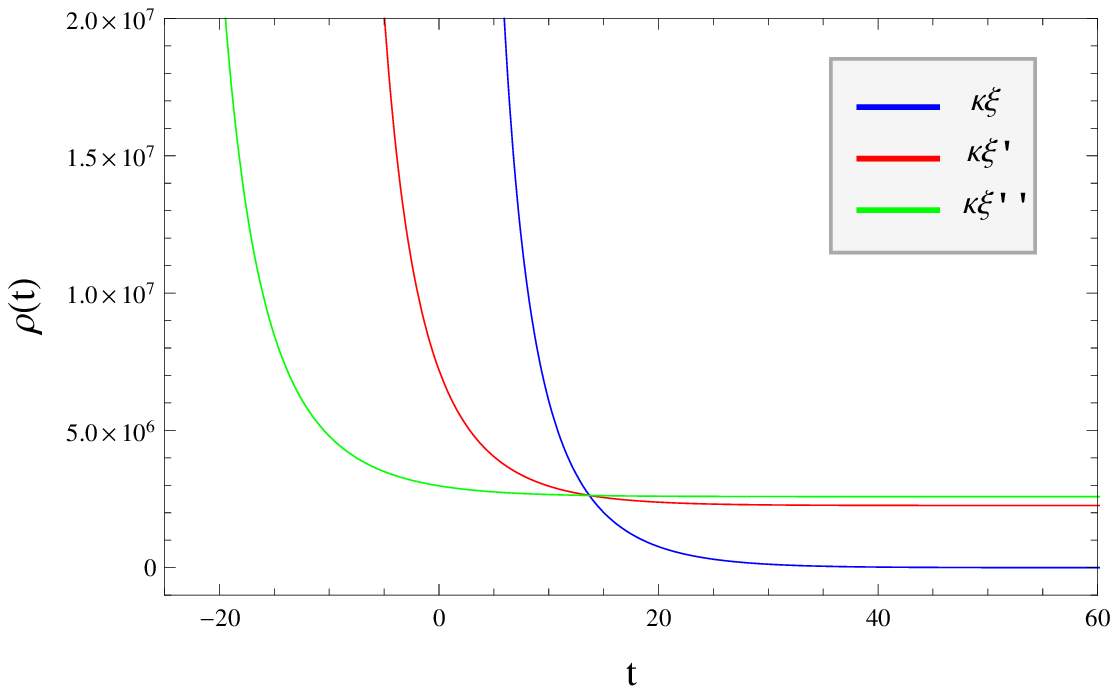}}
  \caption[The scale factor as a function of time for the $\Lambda \textrm{CDM}$ model with and without viscosity]{The time evolution of the scale factor, the Hubble parameter and the energy density for the non-relativistic matter for the $\Lambda \textrm{CDM}$ model with and without viscosity is plotted in Fig.\ref{fig:figc4a}, \ref{fig:fighgenn} and \ref{fig:figrhogenn}, respectively. Here $\kappa\xi:$ $\kappa\xi_{0}=.0$, $\kappa\xi_{1}=.0$, $\kappa\xi_{2}=.0$, $\kappa\xi^{'}:$ $\kappa\xi_{0}=.015$, $\kappa\xi_{1}=.05$, $\kappa\xi_{2}=.01$ and $\kappa\xi^{''}:$ $\kappa\xi_{0}=.01$, $\kappa\xi_{1}=.15$, $\kappa\xi_{2}=.05$. Here we have set $\Omega_{\Lambda 0}= 0.7$, $\Omega_{m 0}= 0.3$, $H_{0} = 0.07$. The time is given in billion years.}
  \label{fig:figrhoh}
\end{figure}

In this case the differential equation for the Hubble parameter and the continuity equation are given in equations (\ref{eq:ni}) and (\ref{eq:aatte}), respectively. Combining these two equations we obtain
\begin{equation}
\kappa\dot{\rho}_{m}=6H\dot{H}.\label{eq:tretti}
\end{equation} 
Integrating eq.(\ref{eq:tretti}), we obtain
\begin{equation}
3H^{2}=\kappa\rho_{m}+\hat{D},\label{eq:trettia}
\end{equation}
where we have defined 
\begin{equation}
\hat{D}\equiv 3H_{0}^{2}-\kappa\rho_{m0}.
\end{equation}
Equation (\ref{eq:trettia}) tells that at the beginning of the cosmic evolution when the density is very large the Hubble parameter is also very large. When the energy density decreases the Hubble parameter will also decrease. We can see this fact from the numerical solutions of the equations (\ref{eq:ni}) and (\ref{eq:aatte}) in Figure \ref{fig:figrhoh}. These Figures show that for this model with viscosity the energy density of the non-relativistic matter and the Hubble parameter diverge at a point of time $t<0$, it means that the universe begins with a big bang. In Figure \ref{fig:figc4a} we have also given the numerical solution for the scale factor for this case. This figure shows that the universe starts with zero volume, and as $t$ increases the scale factor will increase exponentially. 

Figure \ref{fig:figrhoh} also shows that depending on the value of the bulk viscosity coefficients, the universe will expand faster or slower. We also see that as $t\rightarrow \infty$ the energy density will approach a finite value, and for different values of the bulk viscosity coefficients this finite value can be bigger or smaller.    

\begin{figure}
  \centering
  \subfigure[][$q(t)$.]{\label{fig:fig4q1}\includegraphics[width=0.4\textwidth]{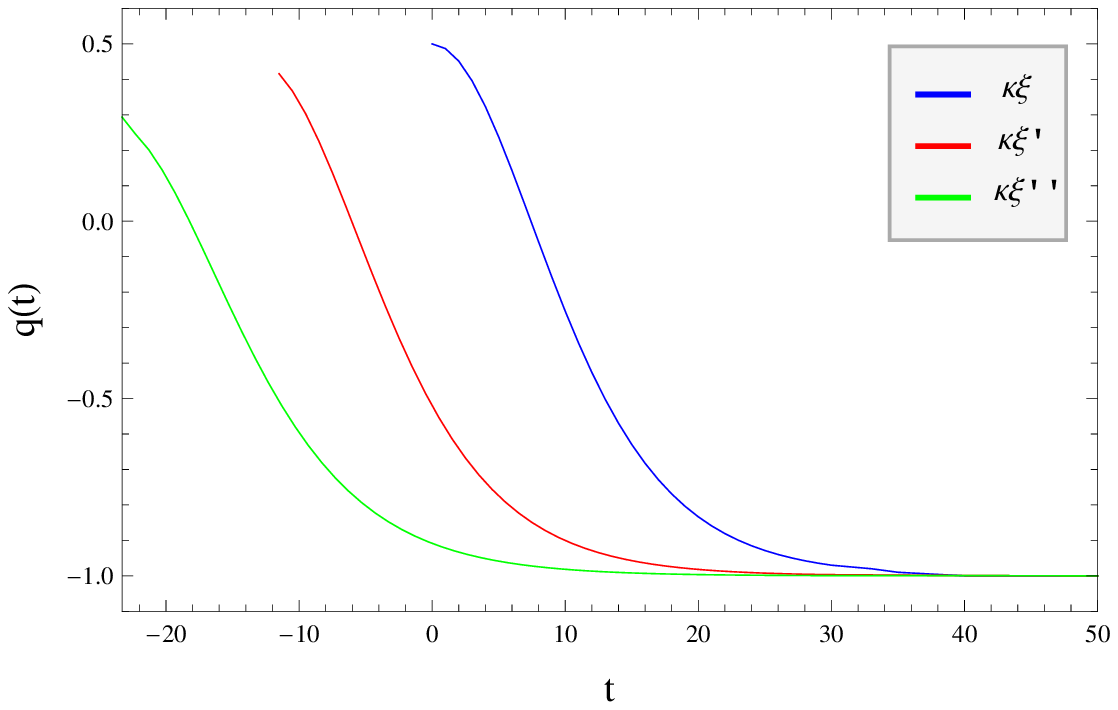}} 
  \subfigure[][$r(t)$.]{\label{fig:fig4cq2}\includegraphics[width=0.4\textwidth]{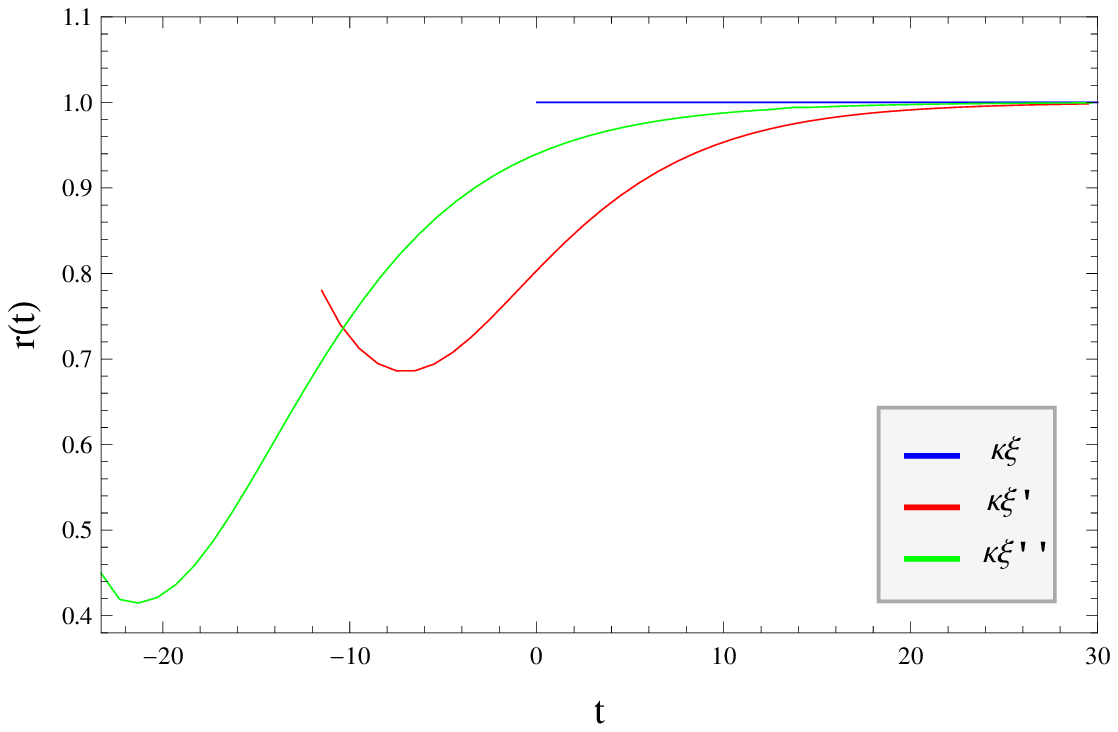}}             \subfigure[][$s(t)$.]{\label{fig:fig4cq3}\includegraphics[width=0.4\textwidth]{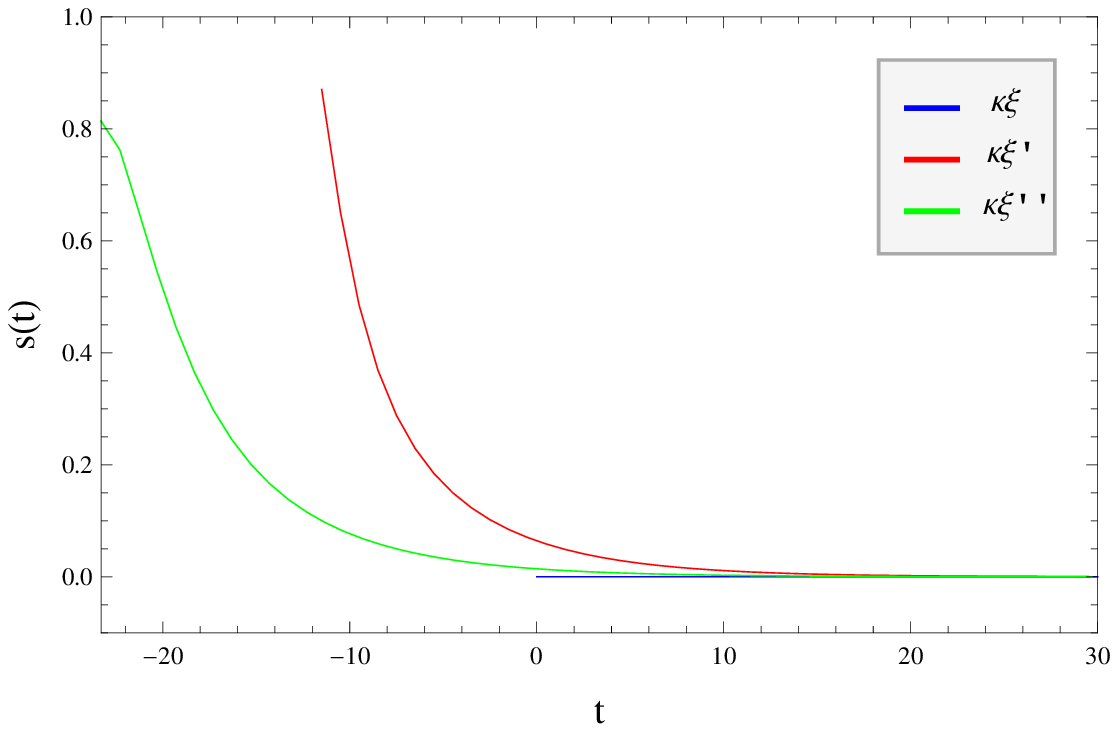}}
  \caption[The scale factor as a function of time for the $\Lambda \textrm{CDM}$ model with and without viscosity]{The time evolution of the deceleration parameter and the statefinder parameters for the $\Lambda \textrm{CDM}$ model with and without viscosity. Here $\kappa\xi:$ $\kappa\xi_{0}=.0$, $\kappa\xi_{1}=.0$, $\kappa\xi_{2}=.0$, $\kappa\xi^{'}:$ $\kappa\xi_{0}=.015$, $\kappa\xi_{1}=.05$, $\kappa\xi_{2}=.001$ and $\kappa\xi^{''}:$ $\kappa\xi_{0}=.01$, $\kappa\xi_{1}=.15$, $\kappa\xi_{2}=.005$. Here we have set $\Omega_{\Lambda 0}= 0.7$, $\Omega_{m 0}= 0.3$, $H_{0} = 0.07$. The time is given in billion years.}
  \label{fig:figq1q2q3}
\end{figure}

In Figure \ref{fig:figq1q2q3} we have plotted the time evolution of the deceleration parameter and the statefinder parameters for different values of the bulk viscosity coefficients. From these figures we can see that the deceleration parameter starts with a positive value at the beginning of the cosmic evolution and as $t$ increases it will approach $-1$. The statefinder parameters will eventually approach the values for the $\Lambda \textrm{CDM}$ model with no viscosity as $t$ increases.

\begin{figure}
  \centering
  \includegraphics[width=0.4\textwidth]{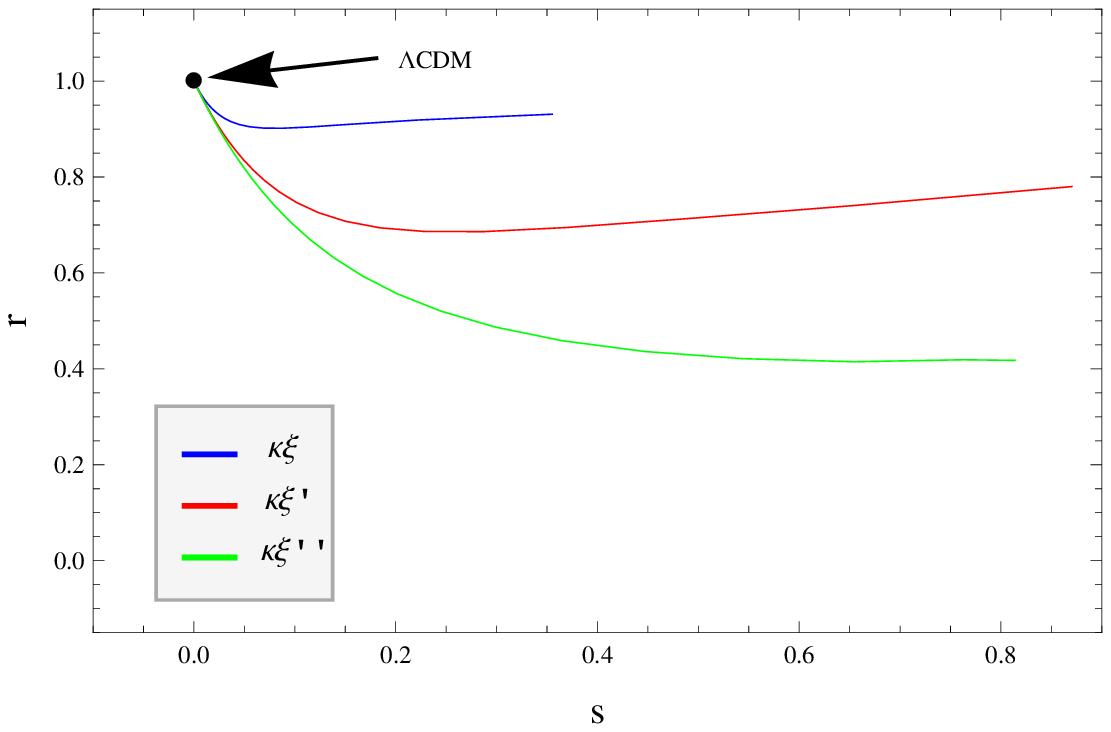}
  \caption[The $s$-$r$-plane for the $\Lambda \textrm{CDM}$ model with and without viscosity]{The $s$-$r$-plane for for the $\Lambda \textrm{CDM}$ model with bulk viscosity. Here $\kappa\xi:$ $\kappa\xi_{0}=.0$, $\kappa\xi_{1}=.0$, $\kappa\xi_{2}=.0$, $\kappa\xi^{'}:$ $\kappa\xi_{0}=.015$, $\kappa\xi_{1}=.05$, $\kappa\xi_{2}=.001$ and $\kappa\xi^{''}:$ $\kappa\xi_{0}=.01$, $\kappa\xi_{1}=.15$, $\kappa\xi_{2}=.005$. Here we have set $\Omega_{\Lambda 0}= 0.7$, $\Omega_{m 0}= 0.3$, $H_{0} = 0.07$.}
  \label{fig:fig4crs}
\end{figure}

In Figure \ref{fig:fig4crs} we have plotted the $s$-$r$-plane for this model, and we see that the curves in this plane will approach the point $\left\{s,r\right\}=\left\{0,1\right\}$, which corresponds to the values for the statefinder parameters for the $\Lambda \textrm{CDM}$ model with no viscosity. 

\section{Results and Conclusion}

In this paper we have studied the $\Lambda \textrm{CDM}$ model with bulk viscosity. We have explored four different cases of bulk viscosity for this model. In the first case the bulk viscosity coefficient was a constant. We found the analytical solutions for the scale factor, the Hubble parameter, the energy density of the non-relativistic matter, the deceleration parameter and the statefinder parameters. In the second case the bulk viscosity was proportional to the Hubble parameter. For this case we gave also the analytical solutions for $a(t)$, $H(t)$, $\rho(t)$, $q(t)$, $r(t)$ and $s(t)$. In both cases we found that due to viscosity the energy density will not approach zero, and that the universe expands faster for bigger values of viscosity. In the third case the bulk viscosity was given as a superposition of the first two cases. For this case we found the analytical solutions of $a(t)$, $H(t)$, $q(t)$, $r(t)$ and $s(t)$, and we gave a numerical solution for the continuity equation. In this case we found that the universe begins with a big bang, but at an earlier point of time, $t<0$. The fourth case of bulk viscosity had the form given in equation (\ref{eq:syv}). In this case we gave the numerical solution of the Raychaudhuri equation and the continuity equation and we found similar results to the third case. We found also that as $t$ increases the statefinder parameters will approach the same values as in the $\Lambda \textrm{CDM}$ model with no viscosity.

From the results of this paper we can conclude that the $\Lambda \textrm{CDM}$ models with viscosity differ from the $\Lambda \textrm{CDM}$ models with no viscosity. For the $\Lambda \textrm{CDM}$ model with viscosity the energy density will not approach zero in the infinitely far future as it is the case for the $\Lambda \textrm{CDM}$ model with no viscosity. From the plots of $s$-$r$-plane we can also conclude that by use of statefinder parameter diagnostic method we can differentiate the viscous $\Lambda\textrm{CDM}$ models from non-viscous cases.

\bibliographystyle{spr-mp-nameyear-cnd}
%

\end{document}